\PassOptionsToPackage{table}{xcolor}
\documentclass[]{aa}  
\usepackage[utf8]{inputenc}
\usepackage{graphicx}
\usepackage{subcaption}
\usepackage{txfonts}
\usepackage{siunitx}
\usepackage{hyperref}
\usepackage{placeins}
\usepackage{multirow}
\usepackage{array}
\usepackage{booktabs}
\usepackage[table]{xcolor}
\usepackage{mathtools}
\usepackage[normalem]{ulem}

\newcommand\redout{\bgroup\markoverwith
{\textcolor{red}{\rule[0.5ex]{2pt}{0.8pt}}}\ULon}

\begin{document} 
   \title{Multimodality for improved CNN photometric redshifts}

\author{  R. Ait Ouahmed \inst{1}\thanks{\email{reda.ait-ouahmed@lam.fr}}
        \and S. Arnouts \inst{1}
        \and J. Pasquet \inst{2,3}
        \and M. Treyer \inst{1}
        \and E. Bertin \inst{4}
}

\institute{ Aix Marseille Univ, CNRS, CNES, LAM, Marseille, France
        \and AMIS - Université Paul-Valéry - Montpellier 3\label{inst2}
        \and UMR TETIS - INRAE, AgroParisTech, Cirad, CNRS, Montpellier, France\label{inst3}
        \and Sorbonne Universit\'e, CNRS, UMR 7095, Institut d’Astrophysique de Paris,      98 bis bd Arago, 75014 Paris, France
} 

   \date{Received date; accepted date}
%%%%%%%%%%%%%%%%%%%%%%%%%%%%%%%%%%%
\abstract{
Photometric redshift estimation plays a crucial role in modern cosmological surveys for studying the universe's large-scale structures and the evolution of galaxies. Deep learning has emerged as a powerful method to produce accurate photometric redshift estimates from multi-band images of galaxies. Here, we introduce a multimodal approach consisting of the parallel processing of several subsets of image bands prior, the outputs of which are then merged for further processing through a convolutional neural network (CNN). We evaluate the performance of our method using three surveys: the Sloan Digital Sky Survey (SDSS), The Canada-France-Hawaii Telescope Legacy Survey (CFHTLS) and Hyper Suprime-Cam (HSC).  
By improving the model's ability to capture information embedded in the correlation between different bands, our technique surpasses the state-of-the-art photometric redshift precision. 
We find that the positive gain does not depend on the specific architecture of the CNN and that it increases with the number of photometric filters available. }
   \maketitle

\section{Introduction}
\label{sec:intro}
Photometric redshifts have become crucial for cosmological surveys based on multi-band imaging surveys such as the current Dark Energy Survey (DES; DES Collaboration et al. 2016), the Kilo-Degree Survey \citep[KIDS, ][]{deJong2013} and the upcoming Vera Rubin survey \citep[LSST, ][]{Ivezic2019} and Euclid \citep[]{Laureijs2011}. 
The magnitude depth and the extent of the area covered by these surveys make it impossible to rely solely on spectroscopy for redshift estimates, so that, photometric redshifts became a major component of these cosmological endeavors.

The methods to estimate redshifts from multi-band photometry fall in three broad  categories: 

\begin{itemize}
\item  Spectral Energy Distribution (SED) template fitting: this technique has been used for several decades. It relies on a set of observed or modeled SEDs, assumed to be  representative of the diversity of galaxies. These theoretical magnitudes are then compared to the observed ones with a minimization fitting procedure to derive the most probable template and redshift estimates \citep[e.g.][]{Arnouts1999,Ilbert2006,Benitez2000,Brammer2008}.

\item Machine Learning algorithms: this approach benefits from the increase in available spectroscopic redshifts required to train the algorithms. Models learn correlations between redshift and the input features provided.
Once trained, they can be used to estimate redshifts based on the same input information. Different algorithms have been used such as artificial neural network \citep[]{Collister2004}, $k-$Nearest Neighbors \citep[kNN,][]{Csabai2007} or random forest techniques \citep[]{carliles2010random}. These methods are fast and were shown to be effective in the domain of validity of the training set. 
As with SED fitting algorithms, the input information consists of features extracted from the multi-band images, such as fluxes, colors, and morphological parameters.   

\item Deep learning algorithms: the images are used directly as input, in contrast to the two previous methods. These algorithms are multiple-layer neural networks that extract relevant features from the multi-band images of galaxies by adjusting parameters during a learning process where a cost function is minimized. 
Convolutional Neural Networks (CNNs) are a popular type of deep learning algorithm for image-related tasks. CNNs are designed to detect small, local correlations and patterns in images with the first layers, and increasingly larger and more complex patterns with deeper layers.

\end{itemize}

Over the last few years, deep learning has proven to be a highly effective method. Through the use of various deep learning frameworks, state of the art results have been achieved in photometric redshift estimation in the Main Galaxy Sample of the Sloan Digital Sky Survey (SDSS MGS), a nearly complete spectroscopic dataset to $r=17.8$. 

 \cite{Pasquet2019} (hereafter P19)  developed a deep CNN based on the Inception network.
 The method uses 64$\times$64 pixel images centered on the SDSS spectroscopic targets in the $ugriz$ bands, along with the line-of-sight galactic reddening value.
 The results show an improvement by a factor of 1.5 over the released SDSS photometric redshifts, which are based on a k-NN algorithm \citep[][]{Beck2016}.

\citet[]{Hayat2021} presented a self-supervised contrastive learning framework. It aims to build a low dimensional space that captures the underlying structure and meaningful features of a large dataset of unlabeled (no spectroscopic redshift) galaxies. The network is trained to minimize the distance between representations of a source image and its augmented versions while maximizing the distance between these representations and representations of other galaxies. Once this latent space is obtained, the network is fine-tuned on the redshift estimation task with labeled data. This work outperforms P19 but more interestingly, it reveals that including unlabeled data reduces the amount of labeled data necessary to achieve the P19 results.

\citet[]{Dey2021} used Deep Capsule Networks to jointly estimate redshift and morphological type. 
Their network consists of a primary convolutional layer followed by Conv-Caps layers. While conventional CNNs primarily detect features, capsule networks also compute feature properties (orientation, size, colors, etc.). Note that even though these networks are robust and invariant to image orientation, the authors used rotation and flip data augmentation during training. The dimension of their latent space is only 16, which allows for a better interpretability. Compared to classical CNNs, the capsule networks are more difficult to train, and not easy to scale to deeper architectures for more complex tasks. Their results on the SDSS sample show a marginal improvement over P19.

Finally, \citet[][submitted]{treyer2023} present an updated version of the network introduced by P19. 
The number of parameters is reduced with a latent space of 96 dimensions instead of 22272 in the original work, which improves the generalization capacity of the network. 
%[leading to better redshift estimation results]. 
While their goal is to extend redshift estimation to fainter magnitude, they also show that the new network outperforms previous works on the SDSS MGS (see Table.~\ref{table:sdss}).

In this work, we propose a multimodal architecture.
Multimodality commonly refers to the combination of different types of information for training \citep[][]{ngiam2011multimodal,ma2015multimodal,hou2018audio}. This approach is especially relevant when dealing with data from different sensors (such as cameras, LiDAR, and radar). It exploits the complementary nature of the information contained in different types of data \citep[e.g., ][]{qian2021robust, chen2017deep}
by processing them in parallel modalities, allowing them  to interact at various stages and finally merging them all together \citep[]{hong2020more}.

The photometric images provide a low-resolution view of the source spectra, and the correlation between them is strongly informative of the redshift. The conventional approach is to  stack these images all together as a network input (P19, \cite{Hayat2021,Dey2021,treyer2023}). In this work, we show that this is suboptimal and we introduce the use of multimodality for redshift estimation. It consists in organizing the input into subsets of bands that are processed in parallel prior to being merged, which improves the extraction of inter-band correlations, and ultimately the redshift precision. Furthermore, we discuss the key ingredients of the multimodal architecture and validate it on several datasets.

%\sout{Here we explore whether organizing the multi-band input into modalities containing subsets of bands can improve the extraction of inter-band correlations, and ultimately the redshift precision. We discuss the key ingredients of the multimodal architecture and optimize them by testing the network on several datasets.}

The paper is organized as follows: the different photometric and spectroscopic datasets are presented in Section 2; the architecture, training and input/output of the network are described in Section 3, with additional information in Appendix A; the adaptation of the network to incorporate multimodality is described in Section 4; Section 5 defines the metrics used to evaluate the redshift estimates and presents different experiments to understand the key components of the multimodal approach; Section 6 presents the performance and gains of the optimal multimodal network with respect to the baseline model (single modality) on different datasets; discussions are made in Section 7 and we conclude this work in Section 8.
\section{Data}
\label{sec:data}
 We use three different photometric and spectroscopic datasets covering a wide range of image depth and redshift. In the following, DR stands for Data Release. 
 \subsection{ The SDSS survey}
The SDSS is a 5-band ($ugriz$) imaging (r$\le$22.5) and spectroscopic survey using a dedicated 2.5-meter telescope at Apache Point Observatory in New Mexico. We use the same spectroscopic sample as P19 based on the SDSS DR12 \citep[][]{Alam2015} in the northern galactic cap and Stripe82 regions. It consists of 516,525 sources with dereddened petrosian magnitudes $r\le 17.8$ and spectroscopic redshifts $z\le 0.4$. For each source, in each of the 5 bands, all the available images from the SDSS Science Archive Server are resampled, stacked and clipped. The resulting input data is a $5\times64\times 64 $ pixel datacube with a pixel scale of 0.396 arcsec, in a gnomonic projection centered on the galaxy coordinates, and aligned with the local celestial coordinate system (see P19 for details), in addition to the galactic extinction value \citep[]{Schlegel1998}. 
%The datacube for the full sample can be retrieved at {\bf XXX}. 
%%
\subsection{The CFHTLS imaging survey}
 The Canada-France-Hawaii Telescope (CFHT) Legacy Survey\footnote{http://www.cfht.hawaii.edu/Science/CFHTLS/}
   (CFHTLS) is an imaging survey performed with MegaCam \citep[]{Boulade2000} in five optical bands ($u^{\star}griz$). 
 In the following we only considered the CFHTLS-Wide component, which covers four independent sky patches totaling 154 deg$^2$ with sub-arcsecond seeing (median$\sim$0.7”) and a typical depth of $i\sim$24.8 (5$\sigma$ detection in 2" apertures).
   
   We use the images and photometric catalogues from the T0007 release\footnote{http://terapix.iap.fr/cplt/T0007/doc/T0007-doc.html} produced by TERAPIX\footnote{http://terapix.iap.fr/} \citep[]{Hudelot2012}. This final release includes an improved absolute and internal photometric calibration, at a 1-2\% level, based on the photometric calibration scheme adopted by the Supernova Legacy Survey \citep[SNLS][]{Regnault2009}. 
  
   The final images are stacked with the Swarp tool\footnote{http://astromatic.net/software/swarp} \citep[]{Bertin2006}. The detection and photometric catalogues were performed with SExtractor \citep[]{Bertin1996} in dual mode with the source detection based on the $gri-\chi^2$ image \citep[]{Szalay1999}. Although the pixel scale is smaller (i.e., 0.18 arcsec/pix) than in the SDSS, we adopt the same $64\times 64$ pixel cutouts for the CFHTLS datacubes.   
\subsection{The HSC-Deep imaging survey}
This dataset consists of the four HSC-Deep fields (COSMOS, XMM-LSS, ELAIS-N1 and DEEP2-3) partially covered by the $u$-band CLAUDS survey \citep[][]{Sawicki2019} and the near-infrared (NIR) surveys UltraVISTA \citep[][COSMOS field]{McCracken2012} and VIDEO \citep[][XMM-LSS field]{Jarvis2013}. A full description of the HSC-Deep dataset and its ancillary data are detailed in \citet[][]{Desprez2023} and summarized hereafter.

The HSC-SSP is an imaging survey conducted with the Hyper Suprime-Cam camera \citep{Miyazaki2018} on the Subaru telescope in 5 broadband filters ($grizy$)\footnote{The HSC-Deep survey include also narrowband filters not considered in this work}. We use the public DR2 \citep[][]{Aihara2019} for the Deep ($\sim$ 20 deg$^2$) and UltraDeep ($\sim$3 deg$^2$) layers of the survey. These have median depths $g=26.5-27$ and $y=24.5-25.5$ respectively.

CLAUDS is a deep survey with the CFHT MegaCam imager in the $u$-band and slightly redder $u^{\star}$-band \citep[][]{Sawicki2019}. The  $u^{\star}$ filter covers the whole XMM-LSS region. ELAIS-N1 and DEEP2-3 are exclusively covered with the 
%new 
$u$ filter, while COSMOS was observed with both filters. CLAUDS covers 18 deg$^{2}$ of the four HSC-Deep fields down to a median depth $u=27$, and 1.6 deg$^2$ of the two ultradeep regions down to $u=27.4$.

UltraVISTA\footnote{\url{https://ultravista.org}} and VIDEO\footnote{\url{http://www.eso.org/sci/observing/phase3/data_releases.html}} are deep NIR surveys acquired by the VISTA Telescope \citep{Emerson2004} with the VIRCAM instrument \citep{Dalton2006}. For UltraVISTA we use the DR3 $YJHK_s$ images covering 1.4~deg$^2$ down to $Y\sim 25$ and $J, H, K_s\sim 24.7$ \citep{McCracken2012}. 

For VIDEO we use the DR4 images in the same passbands covering 4.1~deg$^{2}$, down to depths ranging from $Y=25.0$ to $K_s=23.8$ \citep{Jarvis2013}.

All the images are projected onto the same HSC reference pixel grid, using \textsc{SWarp} \citep{2002ASPC..281..228B}, with a pixel scale of $\ang{;;0.168}$/pixel.  For the $u$-band images, the stacks are  generated with the native HSC pixel grid, while for the NIR images the fully calibrated mosaics are later projected onto the HSC pixel grid.

The dimension of the HSC-Deep datacubes is $9\times64\times 64 $ pixels. They include one $u$-band image ($u^{\star}$, otherwise  $u$), five HSC images ($grizy$) and three NIR images (JHK$_s$). When missing, the NIR channels are padded with zeros.   

\begin{table}[]
\begin{tabular}{|l|c|c|r|}
%%%%%%%%%%%%%%%%%%%%%%%%%%%%%%%%%
\hline  % \noalign{\vskip 0.1cm}
\multicolumn{4}{|c|}{Spectroscopy } \\
\hline % \noalign{\vskip 0.1cm}
Survey             & Res.          &  z-range           & Selection  \\[0.1cm] 
\hline % \noalign{\vskip 0.1cm}
SDSS DR12$^{(1)}$  &  2000         &  $z \le$0.4        & $r\le 17.8$      \\
SDSS-BOSS$^{(2)}$   &  2000        &  0.3$\le z \le$0.7 & LRGs             \\  
GAMA$^{(3)}$         &  1300       &  $z \le$0.7        & $r\le 19.8$      \\  
WIGGLEZ$^{(4)}$      &  1300       &  $z \le$1.2        &  $NUV\le 22.8$   \\  
zCOSMOS$^{(5)}$      &  650        &  $z \le$ 1.2$-$5   & $r\le 22.5-25$  \\  
VANDELS$^{(6)}$      &  650        &  1$\le z \le$6     & $H\le 25$       \\  
UDSz$^{(7)}$         &  650        &  $z\le$4           & $K\le 23$       \\  
DEEP2$^{8)}$        &  6000        &  0.7$\le z\le$1.5  &  $r\le 24$      \\  
VVDS$^{(9)}$         &  230        &  $z\le 1.2-$6      &  $i\le 22.5-24$ \\  
VIPERS$^{(10)}$      &  230       & 0.4$\le z\le$1.5    & $i\le 22.5$     \\  
VUDS$^{(11)}$        &  230       &  2$\le z\le$6       & $K\le 23$       \\  
CLAMATO$^{(12)}$     &  1100      &  2$\le z\le$3.5     &  LBGs           \\  
C3R2$^{(13)}$        &  1100      &   $z\le 4$          & SOM             \\
COSMOS$^{(14)}$      &  multiple  & $z\le 4$            & multiple        \\[0.2cm] 
3DHST$^{(15)}$       &  130       &   $z\le$4           &   H$\le$24      \\
PRIMUS$^{(16)}$      &  40        &  $z\le$0.9          & i$\le$22.5      \\
COSMOS20$^{(17)}$     &  photo-z   &   $z\le$6          & i$\le$26.5     \\[0.1cm] 
\hline 
%%%%%%%%%%%%%%%%%%%%%%
\iffalse
\noalign{\vskip 0.1cm}
\hline
\multicolumn{4}{|c|}{Imaging} \\
\hline % \noalign{\vskip 0.1cm}
Survey & Bands & Depth &  Spectro : Size \\
\hline  % \noalign{\vskip 0.1cm}
%%%
SDSS        & $ugriz$ & $r<22.5$   & 1           : 516500  \\[0.1cm] 
\hline  % \noalign{\vskip 0.1cm}
%%%%%%%%%%%%%%%%%%%%%%%%%%%%%%%%%
            &         &            &  1 $-$ 4    : 46194  \\ 
CFHTLS & $ugriz$ & $i\sim 25$      &   5 $-$ 12  : 82739  \\  
            &         &            &  13 $-$ 14  : 6040   \\[0.1cm]   %   
\hline 
% \noalign{\vskip 0.1cm}
%%%%%%%%%%%%%%%%%%%%%%%%%%%%%%%%% 
     &          &                  &  1 $-$ 4    : 8546   \\ 
     &          &                  &  5 $-$ 12   : 34165   \\ 
HSC  & $ugrizy$ &  $i\sim$26.5     &  13 $-$ 14  : 8702   \\ 
     &    +JHK  &  K$\sim$24.2     &  15         : 2200   \\ 
     &          &                  &  16         : 15000   \\ 
     &          &                  &  17         : 43700 \\[0.1cm] 
\hline
\fi 
%%%%%%%%%%%%%%%%%%%%%%%%
\end{tabular}
\caption{ % Summary of the spectroscopic and imaging surveys. Top part:
 Summary of the spectroscopic surveys with their typical spectral resolution, redshift range and main target selection criteria. Surveys 1 to 14 are used for the spectroscopic training/validation datasets. Survey 15 to 17 are used for test only. 
%. Bottom part:  Main characteristics of the imaging surveys and the size of the matched sources with the different spectroscopic surveys. 
}
\label{table:data}
\end{table}

\subsection{The spectroscopic redshift dataset}
The CFHTLS and HSC-Deep regions have been widely covered by large spectroscopic redshift surveys, including:  
 SDSS-BOSS \citep[DR16, available everywhere,][]{SDSS_DR16},
 GAMA \citep[DR3, $r\le 19.8$,][]{Baldry2018}, 
 WiggleZ \citep[final release, NUV$\le$22.8,][]{Drinkwater2018}, 
 VVDS Wide and Deep \citep[$i\le 22.5$ and $i
\le 24$,][]{LeFevre2013}, 
 VUDS \citep[$i\le 25$,][]{LeFevre2015}, 
 DEEP2 \citep[DR4, $r\le 24$,][]{Newman2013},
 VIPERS \citep[DR2, $i\le 22.5$,][]{Scodeggio2018},   
 VANDELS \citep[DR4, high redshift in XMM-LSS, $H\le 25$,][]{Garilli2021},
 CLAMATO \citep[DR1, high redshift LBGs in COSMOS,][]{KGLee2018},
 UDSz \citep[in XMM-LSS,][]{McLure2013,Bradshaw2013} and zCOSMOS-bright \citep[$i\le 22.5$ in COSMOS,][]{Lilly2007}. We also include the COSMOS team's spectroscopic redshift catalog (M. Salvato, private communication), which consists of several optical and NIR spectroscopic follow-ups of X-ray to far-IR/radio sources, high-redshift star-forming and passive galaxies, as well as poorly represented galaxies in multidimensional color space \citep[C3R2,][]{Masters2019}.
  The Table~\ref{table:data} summarizes the main characteristics of the different spectroscopic surveys considered.

For all the above redshift surveys, we only consider the most secure redshifts, identified with high signal-to-noise and several spectral features (equivalent to flags 3 and 4 in VVDS or VIPERS). For duplicated redshifts, we keep the most secure or randomly pick up one when they have similar flag quality.  \\

  The characteristics of the spectroscopic samples associated to each photometric survey are as follows: 
\begin{itemize}
    \item The SDSS sample includes 516,525 sources with  $r\le 17.8$ and spectroscopic redshifts $z\le 0.4$.
    \item The CFHTLS-Wide sample includes $\sim$108,500 secure redshifts distributed as 34\% with $i\le 19.5$, 57\% with $19.5 \le i \le 22.5$ and 9\% with $22.5 \le i \le 25$. 
    \item The HSC-Deep survey includes $\sim$51,000 redshifts with at least six optical bands ($ugrizy$) and 45\% are brighter than $i\sim 22$ and 10\% fainter than $i\sim 24$. Amongst this sample, $\sim$37,400 sources also have NIR bands ($JHK_s$).
\end{itemize}

 In addition, for the HSC-Deep survey, we also include as test samples the low-resolution spectroscopic redshifts from the 3DHST survey  \citep[based on NIR slitless grism spectroscopy,][]{Skelton2014}, the PRIMUS survey  \citep[based on optical prism multi-objects spectroscopy,][]{Coil2011}, and the 30 band photometric redshifts from COSMOS2020 \citep[][]{Weaver2022}, with the spectral resolution reported in Table~\ref{table:data}.
\begin{itemize}
\item For 3DHST, we use the DRv4.1.5 restricted to  secure grism redshift measurement  \citep{Momcheva2016,Skelton2014}. It contains $\sim4,150$ sources with $H_{AB}\le 24$ located in XMM-LSS and COSMOS.
\item For PRIMUS, we restrict the sample to bright sources (i$_{AB}\le$22.5) at moderate redshift ($z\le 0.9$) with the most secure redshifts \citep[][]{Cool2013}. It contains $\sim 19,500$ sources, located in XMM-LSS, COSMOS and DEEP2-3 fields.
\item For COSMOS2020, we use the 30 band photometric redshifts provided by \citet[]{Weaver2022}, who estimated four different photometric redshifts based on two different multi-band photometric catalogues (using two distinct flux extraction software packages) and two different photometric redshift codes.  We compute the mean and standard deviation of these 4 redshifts, $\bar z$ and $\sigma(z)$, and retain those with $\sigma(z)\le 0.1(1+\bar z)$. 
\end{itemize}

\begin{figure*}
    \centering
    \includegraphics[width=0.9\textwidth]{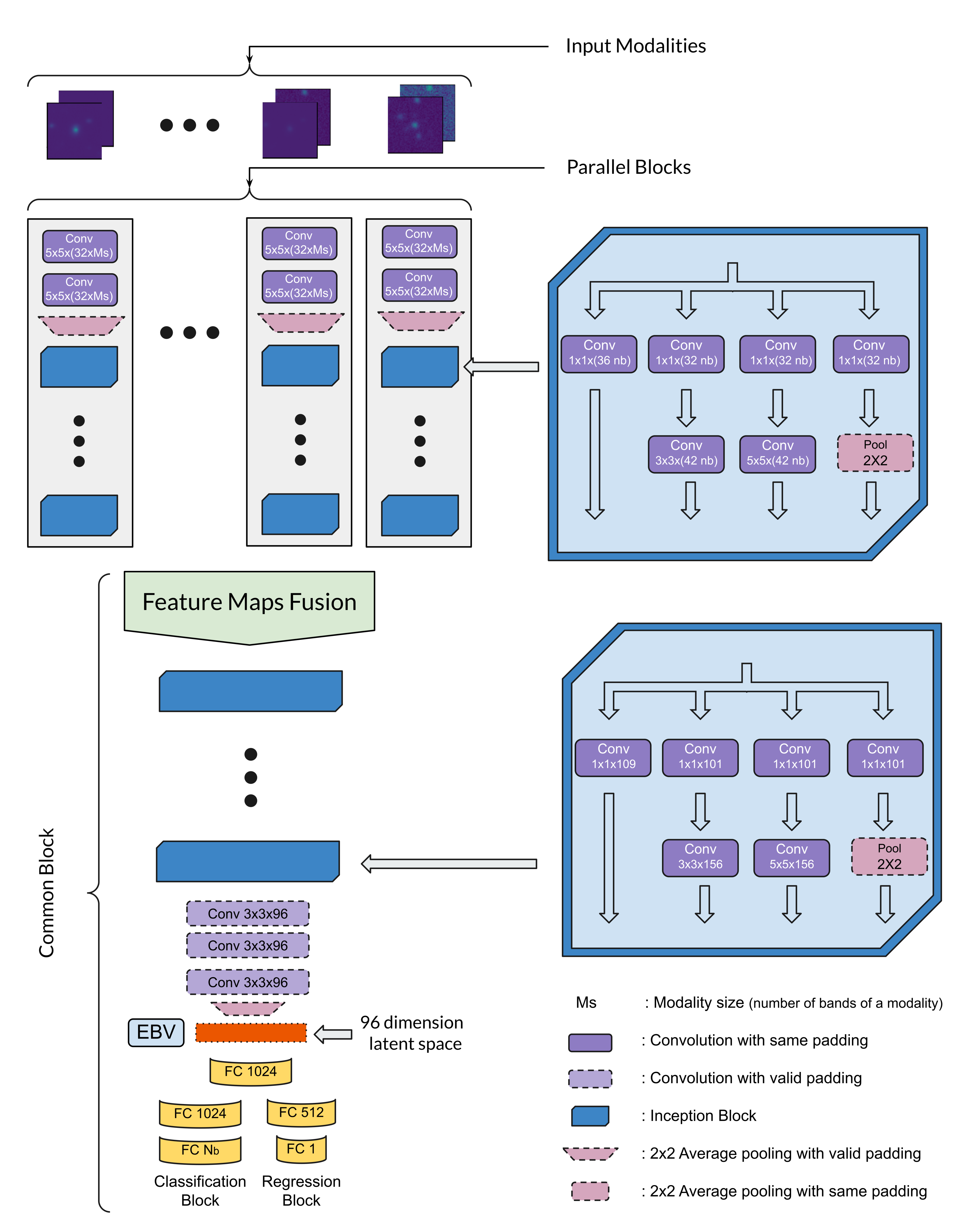}
    \caption{Generic architecture of the multimodality network. The number of parallel blocks is contingent on the number of modalities. The depth of both the parallel and common blocks will be determined by the type of fusion being implemented (early, middle or late fusion), however, it is important to note that the total network depth is fixed at 8 (each modality will go through 8 inception blocks in total). The same goes for the average pooling layers, they are performed, consistently through the different architectures, before the 1st, 4th, and 6th inception blocks and the last one after the valid padding convolution layers. The baseline model without the multi modality approach represents a special case, where all the image bands are grouped into a single modality. The fixed depth allows for a standardized comparison between the different experiments.}
    \label{fig:network}
\end{figure*}
\section{Network and training procedure}
\subsection{Network input}

For each galaxy, a ${\rm N} \times 64 \times 64$ pixel data cube is created with subtracted background. N is the number of bands (5 for SDSS and CFHTLS, 6 or 9 for HSC-Deep). Images in the data cube are sorted in ascending order of wavelength (e.g. $ugriz$).

The network takes as input a batch of datacubes. Given the wide range of pixel values, the P19 dynamic range compression $x_c$ is applied to each image, $x$, defined as $x_c={\rm sign}(x)(\sqrt{|x|+1} -1)$. Additionally, each band is center-reduced using all training objects. This ensures a more robust and efficient training.
 
  %\sout{The network takes as input a batch of datacubes. A dynamic range compression, $x_c$, is applied to each image, $x$, defined as $x_c={\rm sign}(x)(\sqrt{|x|+1} -1)$. Additionally, each band is center-reduced using all training objects.}
  
 Following P19, we also include as input the galactic reddening excess, $E(B-V)$, as the network has no information regarding the location of the sources. The $E(B-V)$ value is appended to the compressed non-spatial latent representation, helping to break the degeneracy between dust reddening and redshift (i.e., P19).

\subsection{The baseline architecture}
\label{subsection:baseline_archi}
As a benchmark, we use a network architecture inspired by P19 and presented by \cite{treyer2023}, which delivers the currently best 
precision on the SDSS MGS dataset (Table.~\ref{table:sdss}). The network consists of two convolutional layers followed by multiple sequential inception blocks \citep[inspired from][]{szegedy2015going}. Each inception block is composed of convolutional layers, with different kernel sizes, which capture patterns at different resolutions. On all layers, a ReLU activation function \citep{nair2010rectified} is used, with the exception of the first and second layers where a PReLU \citep{he2015delving} and a hyperbolic tangent function are employed, respectively, to reduce the signal dynamic range. At the end of the sequential blocks, valid padding is applied, reducing the information to 96 one by one feature maps.
Finally, sequential fully connected layers are employed to produce the classification and regression outputs.

\subsection{Network output}
The redshift estimation task has been treated using either a regression or a classification method. When a regression method is adopted, the network is trained by minimizing a loss function, e.g. the mean absolute error (MAE) or the root mean squared error (RMSE) between the predicted and true redshifts \citep[]{Dey2021,Schuldt2021}. 

Alternatively, it can be treated using a classification method, as in P19 and in this work, and also in other kinds of applications 
\citep{rothe2018deep,stoter2018classification,rogez2017lcr}.
We discretize the redshift space into narrow, mutually exclusive \(N_b\) redshift bins. The network is trained  to classify each galaxy into the correct redshift bin through the optimization of the softmax cross-entropy (a strictly proper loss function). 
 \citet{gneiting2007strictly} show that its correct minimization guarantees convergence to the true conditional probability.

 The outputs of our non-linear, complex enough classification network (after the application of the  softmax activation function) are positive and normalized scores distributed over the predefined redshift bins. We consider them estimators of the true conditional probability of the redshift belonging to a specific bin

\citep{lecun2015deep,krizhevsky2017imagenet,szegedy2015going}, which is, in turn, an approximation of the true redshift probability density function.
Consequently, we will refer to the network classification output as a redshift probablity distribution (PDF). 

In Appendix~\ref{App:ClassvsReg}, we show the performance obtained with models based on  regression and/or classification methods using two different training sets. We find that the classification model outperforms the regression model. Additionally, we obtain a slight improvement by combining the classification and regression losses. In all subsequent experiments, we adopt this mixed scheme.

\subsection{Network Training Protocol}

We use an ADAM optimizer \citep{kingma2014adam} and a batch size of 32 datacubes to train our network.  
Data augmentation is applied with random flips and rotations of the images (90$^{\circ}$ step).

The models are trained by simultaneously optimizing the cross-entropy loss function for the classification module and the MAE for the regression  module. For a source \(s\) with spectroscopic redshift $z_{spec}$, the loss function is the sum of these two loss functions: 
    \begin{equation}
    L(s) =  \sum_{i=1}^{N_b} - y_{i}\log(p_{i}) + |z_{pred} - z_{spec}|
    \label{eq:global_loss}
    \end{equation}
where \(N_b\) is the number of redshift classes, \(y_{i}\) the classification label of the redshift bin \(i\) (1 for the bin containing $z_{spec}$, 0 for the other bins), \(p_{i}\) the estimation for the class $i$ produced by classification module and \(z_{pred}\) the regression  estimate.

For a given training set, the database is split into 5 cross-validation samples. Each cross-validation sample (20\%) is used as a test sample, while the remaining 4 (80\%) are used for training. This guarantees that each galaxy appears once in the test sample. We use ensemble learning \citep{goodfellow2016deep} by running each training configuration several times with weights randomly initialized and the training set randomly shuffled: three times for the HSC and CFHTLS datasets and five times for the SDSS dataset (for comparison with other published works). The final PDF is the average of the outputs of the trained models. 
% chap 7 if I add goodfellow

All the results presented in the following sections are limited to $i\le 24$.

\section{Multimodality for redshift estimation}

A key component of redshift estimation is the correlations between different bands covering different spectral domains. SED fitting techniques and machine learning algorithms exploit the flux ratios between bands. CNNs are able to capture correlations between different channels directly from the images and to extract spatially correlated patterns. In a classical CNN architecture, each kernel of the first convolution layer combines all the channels to produce one feature map (see Fig 3 in P19). 

Multimodality is commonly used to train a network with multiple kinds of input data (i.e. images, audio, text) \citep{ngiam2011multimodal,hou2018audio}. Multiple input streams are incorporated into the network, processed in parallel and combined at a later stage \citep{hong2020more}. This allows for better feature extraction from each modality. 

In the present work, we use multimodality to analyze subsets of bands separately before combining their outputs. In the following, we introduce our formalism for the multimodal configuration, the modifications to the network architecture and the key hyper-parameters involved in such networks.      

\subsection{Modalities}

 The images are sorted in ascending order of wavelength. The size of a modality refers to the number of bands it contains, while the order refers to the proximity of the bands. First-order modalities use adjacent bands, second-order modalities use bands separated by one band, third-order modalities use bands with a gap of two bands, etc. Table.~\ref{table:modalities_order_details} details modalities of first, second and third order for the $ugrizyjhk$ bands.

\subsection{Network architecture}
We adopt a flexible network architecture to incorporate the multimodalities. As illustrated in Fig.~\ref{fig:network}, we define two main parts:

\begin{itemize}
    \item Parallel blocks : for each input modality, we define an independent module at the start of the network. It consists of successive inception blocks sized according to the size of the modality. 

    \item Common block: it combines the outputs of the parallel blocks and proceeds with its own architecture detailed in Fig~\ref{fig:network}.     
\end{itemize}

The depth of the parallel and common blocks depend on the type of fusion used as described in the next subsection. However we limit the total network depth to 8 inception blocks and the pooling layers are performed at fixed depths (before the 1st, 4th, and 6th inception blocks).
The baseline architecture presented in section \ref{subsection:baseline_archi} can be obtained within the current framework by considering only one modality containing all the bands. In the following we use  "baseline" and "baseline single modality" interchangeably.
%\sout{The baseline model is a special case where all input bands are grouped into a single modality.}

\begin{figure*}[ht!]
    \centering
    \begin{subfigure}[b]{0.45\textwidth}
        \centering
        \includegraphics[width=\textwidth]{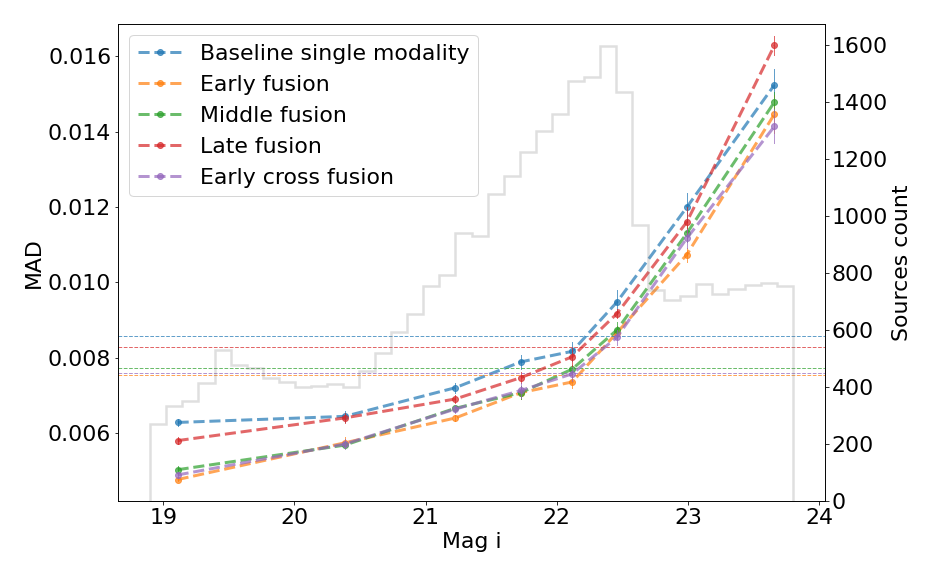}
    \end{subfigure}
    \hspace{\fill}
    \begin{subfigure}[b]{0.45\textwidth}
        \centering
        \includegraphics[width=\textwidth]{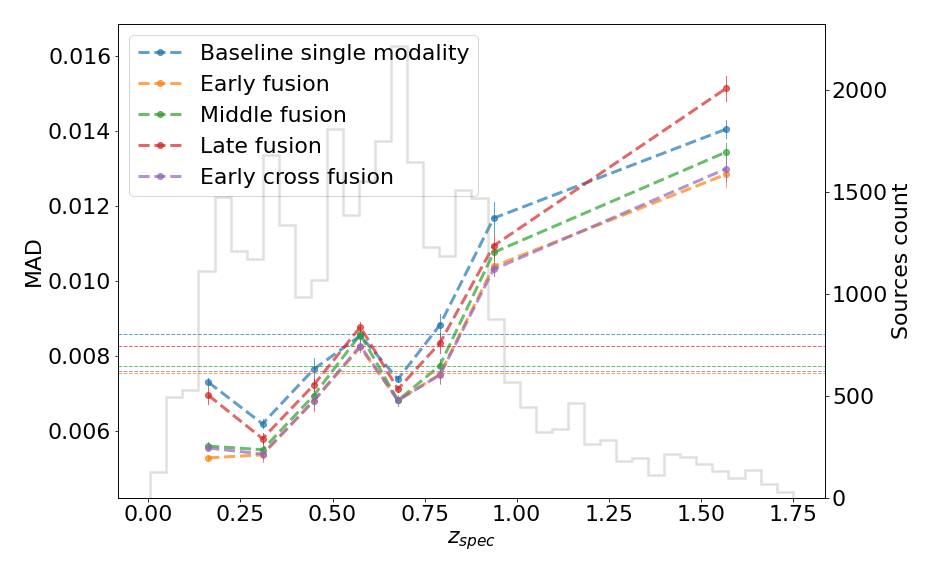}
    \end{subfigure}
    \caption{MAD of the redshift estimation as a function of magnitude (i band, left panel) and spectroscopic redshift (right panel) for different types of fusion, compared to the baseline (single modality) model on the HSC 9-band dataset. The grey histograms represent the magnitude and redshift distributions, the horizontal lines show the mean MAD, and the error bars represent the standard deviation between the 5 validation folds. The data are split into 8 x-axis bins containing the same number of objects, each point representing the center of the bin.}

\label{fig:fusion}
\end{figure*}

\begin{figure*}[ht!]
    \centering
    \begin{subfigure}[b]{0.45\textwidth}
        \centering
        \includegraphics[width=\textwidth]{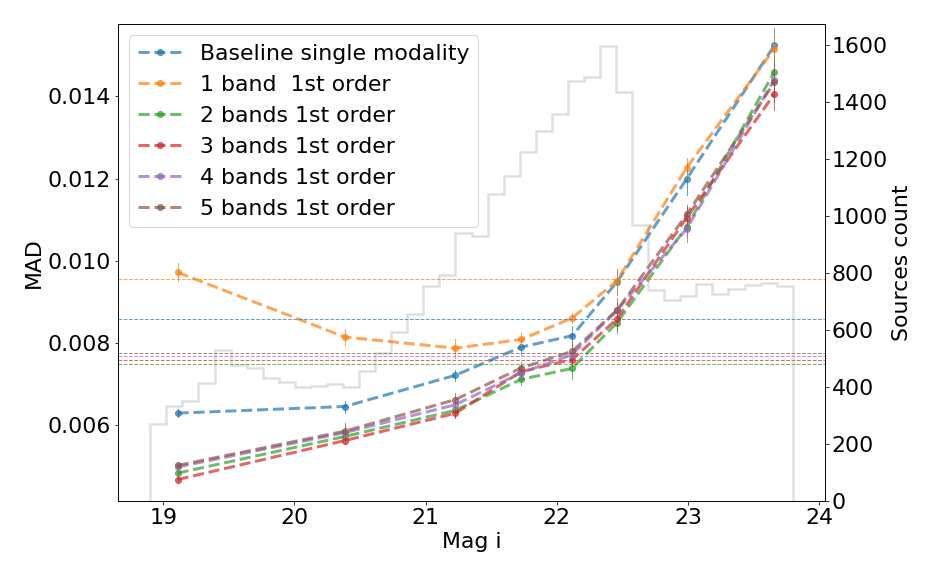}
        
    \end{subfigure}
    \hspace{\fill}
    \begin{subfigure}[b]{0.45\textwidth}
        \centering
        \includegraphics[width=\textwidth]{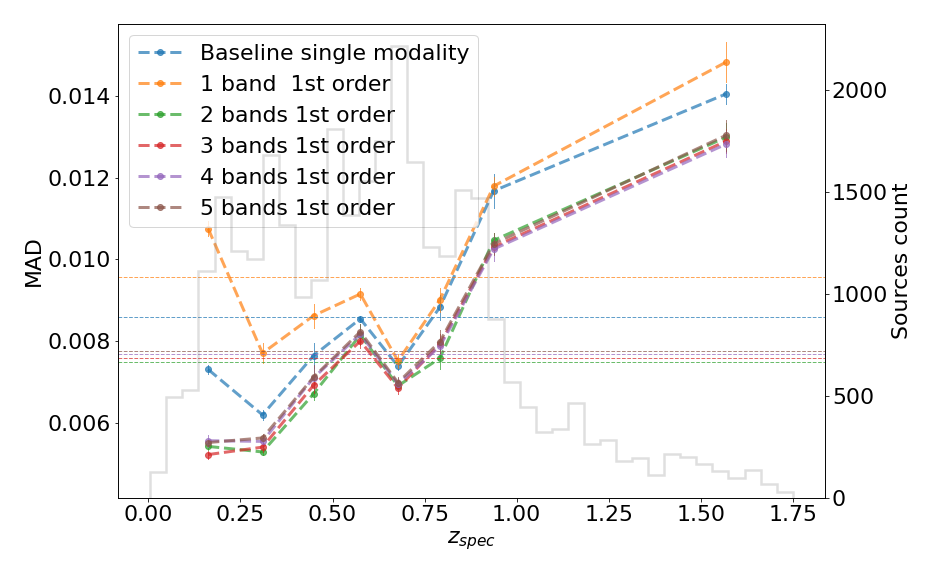}
        
    \end{subfigure}
     \caption{
     %Comparison of the redshift estimation metrics versus magnitude of band i and spectroscopic redshift 
     Same as Fig~\ref{fig:fusion} for early fusion, first order modality models with different number of bands per modality, compared to the baseline model. % on the 9-band sources from the HSC dataset
     }
\label{fig:n_bands_exp}
\end{figure*}

\subsection{Fusion}
The stage of fusion, at which the parallel processed modalities are combined, is the last key factor to consider.  
It determines how much network processing is allocated to feature extraction from each modality and how much is assigned to  combining those features for redshift estimation. We consider the following three stages \citep{hong2020more}:

\begin{itemize}
\item Early fusion: the features from each modality are fused after two parallel inception blocks, prior to passing through six common inception blocks.

\item Middle fusion: modalities are combined after four parallel inception blocks, followed by four common blocks.

\item  Late fusion: modalities are combined after six inception blocks, followed by two common blocks.

\end{itemize}

We test two methods to fuse the feature maps from the different modalities:
simple concatenation and cross-fusion. A cross-fusion module consists in a set of parallel inception blocks, each processing modalities one by one (hence cross) for improved feature blending. 
The cross-fused feature maps pass through a common convolution layer prior to being concatenated \citep{hong2020more}.

%%%%%%%%%%%%%%%%%%%%%%%%%%%%%%%%%
%
%%%%%%%%%%%%%%%%%%%%%%%%%%%%%%%%%%%%%%%%%%%%%%%%%%

\section{Experiments}
\subsection{Metrics and point estimates}
\label{sec:photoz-metrics}
To evaluate the photometric redshift performance between the different experiments, three metrics are considered  based on the normalized residuals $\Delta z= (z_{\rm phot}-z_{\rm spec})/(1+z_{\rm spec})$ (P19):
\begin{itemize}
\item the \textbf{MAD} (Median Absolute Deviation), 
$\sigma_{\rm MAD}=1.4826\times \textrm{Median} ( |\Delta z - \textrm{Median}(\Delta z)|)$
\item the fraction of \textbf{outliers}, $\eta$ (\%), with $|\Delta z|\ge 0.05$ for the SDSS or 0.15 for the other datasets.
\item the \textbf{Bias}, 
$<\Delta z> = \textrm{Mean}(\Delta z)$
\end{itemize}
We choose as point estimate, $z_{phot}$, the median of the output PDF. However this choice is not critical as we are interested in the relative performance of the various experiments. 

\begin{figure*}[ht!]
    \centering
    \begin{subfigure}[b]{0.45\textwidth}
        \centering
        \includegraphics[width=\textwidth]{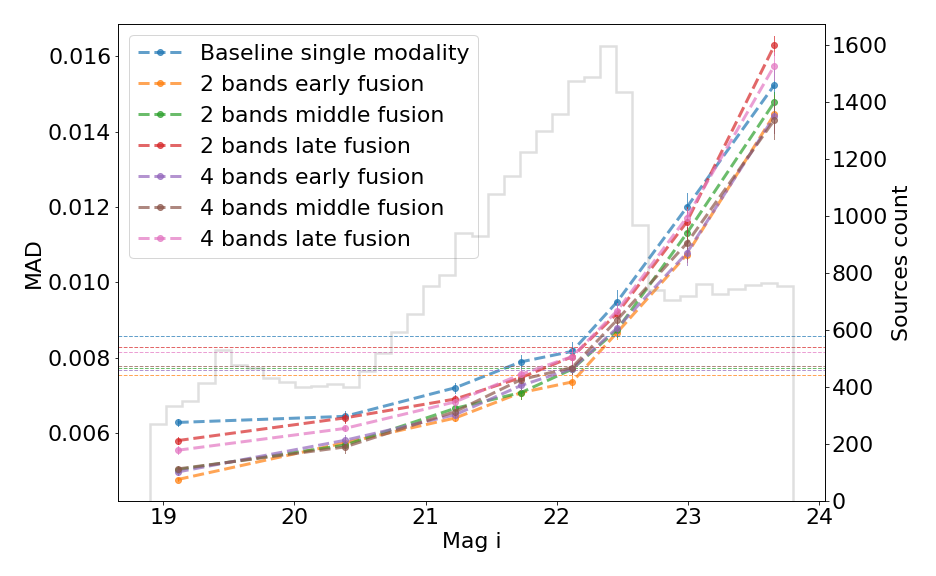}
        
    \end{subfigure}
    \hspace{\fill}
    \begin{subfigure}[b]{0.45\textwidth}
        \centering
        \includegraphics[width=\textwidth]{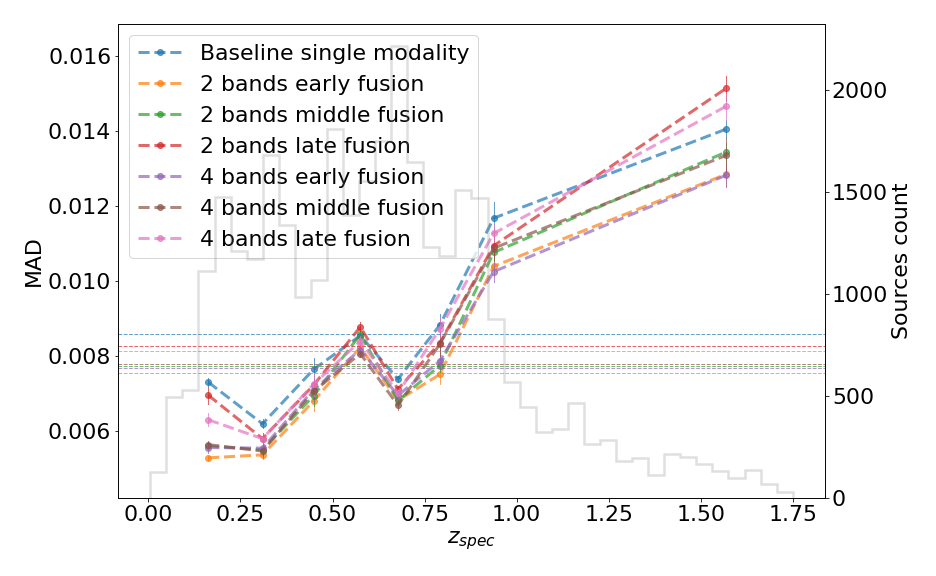}
        
    \end{subfigure}
     \caption{
     %Comparison of the redshift estimation metrics versus magnitude of band i and spectroscopic redshift  
     Same as Fig~\ref{fig:fusion} for modalities of size 2 and 4 using the three different stages of fusion, compared to the baseline model. 
     %on the 9-band sources from the HSC dataset
     }
\label{fig:n_bands_X_fusion_stage_exp}
\end{figure*}

\begin{figure*}[ht!]
    \centering
    \begin{subfigure}[b]{0.45\textwidth}
        \centering
        \includegraphics[width=\textwidth]{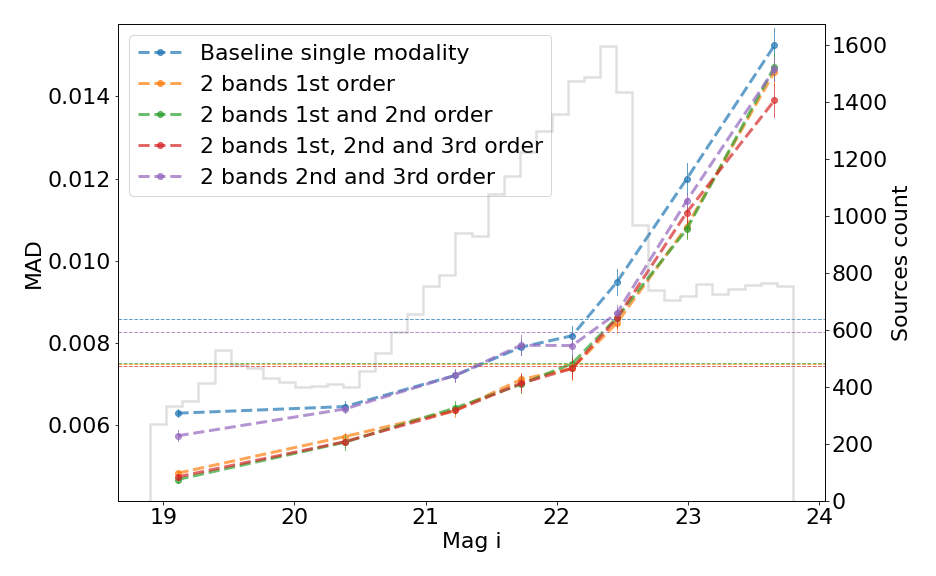}
        
    \end{subfigure}
    \hspace{\fill}
    \begin{subfigure}[b]{0.45\textwidth}
        \centering
        \includegraphics[width=\textwidth]{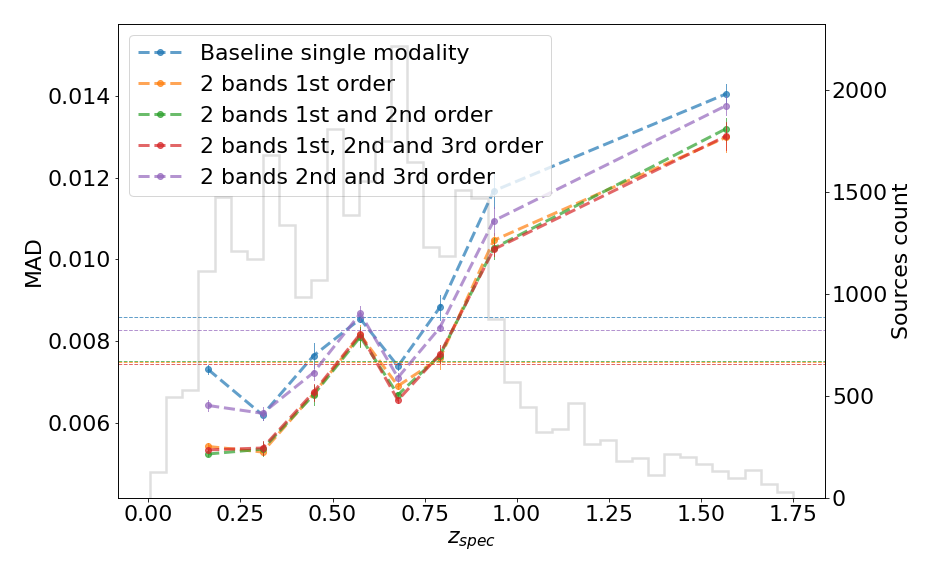}
        
    \end{subfigure}
     \caption{
     %Comparison of the redshift estimation metrics versus magnitude of band i and spectroscopic redshift 
     Same as Fig~\ref{fig:fusion} for different combinations of orders within 2-band modalities and using early fusion, compared to the baseline model. %on the 9-band sources from the HSC dataset
     }
\label{fig:modalities_order}
\end{figure*}
\subsection{Multimodality configurations}

To evaluate the impact of the three variable ingredients of our multimodal approach, we use the HSC Deep Imaging Survey dataset (Section~\ref{sec:data}.3), as it covers the widest range of magnitude and redshift and has the largest number of photometric bands. We run experiments with different multimodal configurations in order to determine :

\begin{itemize}
    
    \item the most efficient stage of fusion and the best fusion type between cross-fusion and simple concatenation,
    \item the optimal modality size,
    \item the optimal modality order, 
\end{itemize}
for extracting redshift relevant information.

\subsubsection{Stage and type of fusion}
We conduct four experiments: early, middle, and late fusion with concatenation fusion, and an early cross fusion scenario,  assuming size 2 and first  and second order modalities.

The resulting  MAD as a function of magnitude and redshift are shown in Fig.~\ref{fig:fusion} and compared to the baseline, single modality model. Error bars are defined as the standard deviation between the metrics of the 5 validation folds. 
Early and middle fusions provide the most significant improvement with early fusion slightly outperforming middle fusion. 
The performance of the early concatenation fusion is similar to the early cross fusion scheme while being more computationally efficient. Thus we will proceed with concatenation fusion for the other experiments.

Additionally, we test very early fusion (where fusion occurs after the two initial convolutions and one inception block) and extremely early fusion ( fusion after just two convolutions). Results reported in Fig~\ref{fig:early_fusion_types} show that early fusion obtains the best precision followed by very early fusion then extremely early fusion. 

\subsubsection{Size of modalities}

 We vary the size of the modalities from one to five assuming first order and early fusion. The MAD are presented in Fig.~\ref{fig:n_bands_exp} as a function of magnitude and redshift.  Adopting a size of 2 or more \textit{significantly and} similarly improves the  performance over the baseline.  This confirms our initial hypothesis that processing subsets of bands in parallel prior to merging information helps the network to capture inter-band correlations. 
 In contrast single band modalities perform similarly to the baseline at faint magnitude and worse at bright magnitude. The network may have more difficulties extracting inter-band correlation information, in this case not available until the modalities are merged within the network. 
 
To further investigate these results, we analyze the impact of modalities of size 2 and 4 under early, middle and late fusion as shown in Fig.~\ref{fig:n_bands_X_fusion_stage_exp}. We can observe the relatively minor impact of the modality size under the three different configurations. 
We conclude that the impact of modality size does not depend on the stage of fusion.

\subsubsection{Order of modalities }

Here we examine the impact of modalities based on the wavelength closeness of their bands. Assuming two-band modalities and early fusion, we test four combinations of orders : first order; first and second; first, second and third order; and finally second and third order (the different orders are detailed in Table~\ref{table:modalities_order_details}).

As illustrated in Fig.~\ref{fig:modalities_order}, we find that 
experiments that included first order modalities performed optimally. The experiment using only second and third order was comparable to the baseline, showing that the network was not able to extract additional relevant inter-band correlation information 
that could outperform the baseline. These results are in line with expectations, as adjacent bands express with the highest resolution the color information directly related to redshift estimation.

\section{Results}
\label{sec:results}
\begin{table}[]
\def\arraystretch{1.3}%
\begin{tabular}{|c|c|c|c|}
\hline
Experiences &  $\sigma$  & $\eta$ & $<\Delta z>$ \\
 &  \num{e-3} & \% & $10^{-3}$ \\
\hline
\multicolumn{4}{|c|}{SDSS r < 17.8}             \\ \hline
 P19 & 9.08 & 0.31  & {\bf0.04}           \\ \hline
 \citet[]{Dey2021} & 8.98 & 0.19  & 0.07           \\ \hline
\citet[]{Hayat2021}  & 8.25 & 0.21  & 0.1         \\ \hline
\cite{treyer2023}  & 8.00 & 0.18  & -0.31          \\ \hline
% Multimodal Network & 07.87 & 0.16 & -0.27              \\ \hline
Multimodal Network & {\bf7.85} & {\bf0.16} & 0.31     \\ \hline
\end{tabular}
\caption{ Performance comparison of different deep learning networks on the SDSS MGS ($r 	\le 17.8$) 
}
\label{table:sdss}
\end{table}

Based on the above experiments, we evaluate the multimodal approach using two-band, first order modalities and performed cross-validations on different datasets.

The SDSS MGS dataset $(r \le 17.8)$ provides a 
%suitable 
benchmark to compare our work with other deep learning redshift estimates \citep[P19;][]{Dey2021,Hayat2021} and with the baseline model \citep{treyer2023}. Results reported in Table.~\ref{table:sdss} show that the multimodal approach outperforms all previous works both in term of MAD and outlier fraction while not worsening the baseline bias.

We compare the multimodal network with the baseline on the  CFHTLS (5 bands), and HSC (6 and 9 bands). Additionally, we test the network trained on the HSC 9 bands on the low resolution spectroscopic samples 3DHST and PRIMUS and on the high quality photometric redshift COSMOS2020. The metrics are reported in Table.~\ref{table:HSC}. We also report the relative gain/loss defined as follows : 

\begin{equation}
G(M)= \frac{|M_B| - |M_M|}{|M_B|}
\end{equation}
 where $M_B$ and $M_M$ are respectively the baseline and multimodality values of a given metric $M$. Finally, we estimate the  statistical significance  of the differences in metrics ($M_B - M_M$) using the paired bootstrap test detailed in Appendix~\ref{paired_bootstrap_test}. 
The computed $p_{values}$ are reported in Table.~\ref{table:HSC} with statistically significant differences under a 5\% risk threshold highlighted in green.  

 The results in Table~\ref{table:HSC} show that the multimodal approach offers statistically significant improvements of the MAD, ranging from 2\% to 10\%,  across all datasets. In the case of 3DHST, the difference is significant under a 7\% risk threshold.
 Similar improvements are also observed in the outlier fractions, ranging from 4\% to 30\%. However, the improvements in the HSC 9 and 6 bands and the 3DHST datasets were not statistically significant under a 5\% risk threshold. Regarding the bias, the baseline approach performs better on the HSC 9 bands and CFHTLS,  but with no significant difference. The  two-band, first order setting achieves these results while being only  1.2 times slower than the baseline.
 
%\sout{ , with a significant difference only for CFHTLS. }

\begin{figure}
\centering
\includegraphics[width=\linewidth]{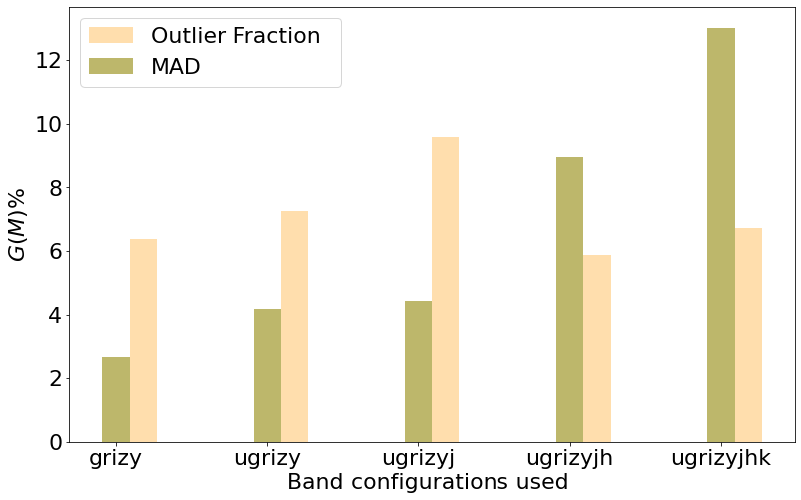}
\caption {Comparison of the multimodality gain  $G(M)$ using 5, 6, 7, 8, and 9 bands  for the MAD and outlier fraction on the HSC 9 band dataset}
\label{fig:Gain_mad_outliers_hsc_9_bands}
\end{figure}

We investigate the relation between the impact multimodality and the number of bands. Figure \ref{fig:Gain_mad_outliers_hsc_9_bands} illustrates the multimodality  gains compared to the baseline when training the models with different band combinations, specifically $grizy$, $ugrizy$, $ugrizyj$, $ugrizyjh$, and $ugrizyjhk$, using the HSC 9 band subset.  We can see that the impact of multimodality on the MAD becomes more pronounced as more bands are incorporated into the training. 

Its effect on the outlier fraction is less conclusive, as it does not exhibit a consistent pattern with the increasing number of bands.

In conclusion, our experiments show that the multimodality approach offers statistically significant improvement in the precision of the redshift estimation. This is observed in both the MAD and the outlier fraction across all datasets. The impact is less conclusive for the mean bias. 

%\sout{In conclusion, our experiments show that adding multimodality to the neural network significantly improves the precision of the redshift estimation. The improvements are observed in both the MAD and the outlier fraction across all datasets and are less conclusive for the mean bias. }

\begin{table}[]
\begin{tabular}{|c|c|c|c|c|}
\hline
Experiences &  $\sigma$  & $\eta$ & $<\Delta z>$  & Count \\
 &  \num{e-3} & \% & $10^{-3}$  &  $10^{3}$ \\
\hline

\multicolumn{5}{|c|}{SDSS}             \\ \hline
Baseline &  07.99 & 0.18  & 0.34  & 516.5          \\ \hline
Multimodal & {\bf 07.85} & {\bf 0.16}  & {\bf 0.31}  & 516.5         \\ \hline
$G(M)$ & 1.74\% & 10.88\%  & 6.28\%  & -         \\ \hline
$p_{value}$ & \cellcolor{green!35} 0.0 & \cellcolor{green!35} 0.0  & \cellcolor{green!35} 0.0  & -         \\ \hline

\multicolumn{5}{|c|}{CFHTLS}             \\ \hline
Baseline & 16.01 & 0.85  & {\bf 0.22}  & 108.5          \\ \hline
Multimodal & {\bf 15.35} & {\bf 0.79}  & 0.29  & 108.5         \\ \hline
$G(M)$ & 4.13\% & 7.22\%  & -24.05\%  & -         \\ \hline
$p_{value}$ & \cellcolor{green!35} 0.0 & \cellcolor{green!35} 0.0002  &  0.15  & -         \\ \hline

\multicolumn{5}{|c|}{ HSC-6b }             \\ \hline
Baseline & 09.14 & 1.25  & 1.97     & 46.8      \\ \hline
Multimodal & {\bf 08.87} & {\bf 1.20}  & {\bf 1.63}  & 46.8       \\ \hline
$G(M)$ & 2.96\% & 3.94\%  & 17.33\%  & -         \\ \hline
$p_{value}$ & \cellcolor{green!35} 0.0 & 0.0575  & \cellcolor{green!35} 0.04  & -         \\ \hline

\multicolumn{5}{|c|}{ HSC-9b }             \\ \hline
Baseline & 08.41 &  1.24  & {\bf 1.58}     & 33.1      \\ \hline
Multimodal & {\bf 07.60} & {\bf 1.19}  & 1.64   & 33.1       \\ \hline
$G(M)$ & 10.1\% & 3.67\%  & -3.1\%  & -         \\ \hline
$p_{value}$ & \cellcolor{green!35} 0.0 & 0.11  & 0.40  & -         \\ \hline

\multicolumn{5}{|c|}{ HSC-9b with 3DHST redshifts}             \\ \hline
Baseline &   14.44 &  2.46  &  13.28  & 2.2         \\ \hline
Multimodal & {\bf 13.88} & {\bf2.37}  & {\bf10.6}   &  2.2       \\ \hline
$G(M)$ & 3.93\% & 3.71\%  & 20.19\%  & -         \\ \hline
$p_{value}$ & 0.069 & 0.27  & 0.10  & -         \\ \hline

\multicolumn{5}{|c|}{ HSC-9b with PRIMUS redshifts}             \\ \hline
Baseline & 12.34 & 2.66  & 11.84   & 15       \\ \hline
Multimodal & {\bf 11.38} & {\bf 1.85}  & {\bf 09.23}    &  15       \\ \hline
$G(M)$ & 7.74\% & 30.4\%  & 22.01\%  & -         \\ \hline
$p_{value}$ & \cellcolor{green!35} 0.0 & \cellcolor{green!35} 0.0  & \cellcolor{green!35} 0.0  & -         \\ \hline

\multicolumn{5}{|c|}{HSC-9b with COSMOS2020 photometric redshifts}             \\ \hline
Baseline & 12.01 & 1.01  & 8.74    & 43.7       \\ \hline
Multimodal & {\bf 11.46} & {\bf 0.83}  & {\bf 6.82}  & 43.7         \\ \hline
$G(M)$ & 4.57\% & 17.08\%  & 21.97\%  & -         \\ \hline
$p_{value}$ & \cellcolor{green!35} 0.0 & \cellcolor{green!35} 0.0  & \cellcolor{green!35} 0.0001  & -         \\ \hline

\end{tabular}
\caption{Impact of multimodality on different datasets. The  MAD, the outlier fraction and the bias are reported for the baseline and the multimodal models, alongside with the relative difference and the $p_{value}$  as a measure of significativity of the observed difference.  The sizes of the datasets {\bf down to $i=24$} are reported on the last column. For the 9 band experiments, some objects had the j band missing, so we used redshift estimations of models trained in those conditions. 
}
\label{table:HSC}
\end{table}

\section{Discussions}
% \subsection{Multimodality impact on different networks}
\subsection{Dependence on network architecture}

\begin{figure}
\centering
\includegraphics[width=\linewidth]{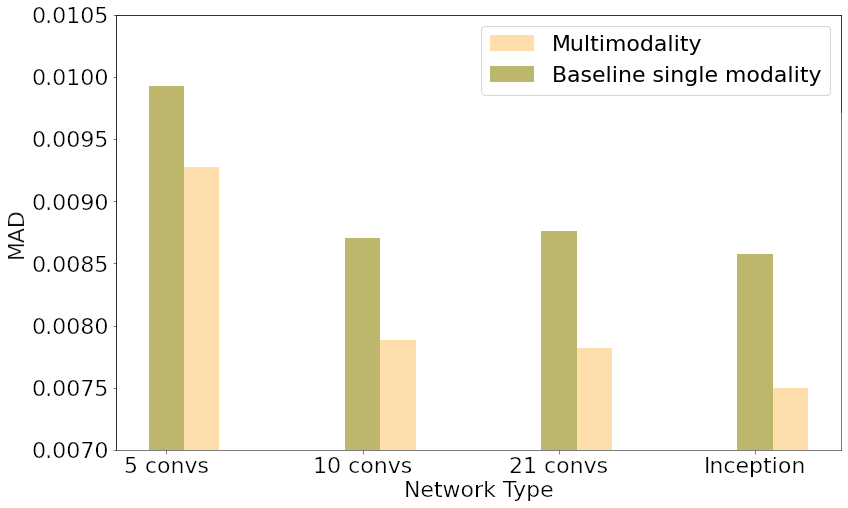}

\caption {Comparison of the multimodality impact on the MAD of the redshift estimation in the HSC 9-band dataset for 4 different network architectures.}
\label{fig:impact_on_different_archis}
\end{figure}

We evaluate the integration of multi-modality in three additional network architectures: a 5-layer CNN, a 10-layer CNN, and a 21-layer CNN. The impact of multimodality on the MAD of redshift estimates for these architectures, as well as the inception baseline, is depicted in Fig.~\ref{fig:impact_on_different_archis}. The results show a consistent improvement when multimodality is incorporated. Its impact was more substantial in the deeper networks compared to the shallower  5 convolution layer network. 
We conclude that the effectiveness of multimodality is enhanced when the network architecture is sufficiently deep. Finally, we  note that these results are unrelated to the number of network parameters as shown in Appendix~\ref{appendix:configuration_details}.

\subsection{Dependence on training set size}

\begin{figure}
\centering
\includegraphics[width=\linewidth]{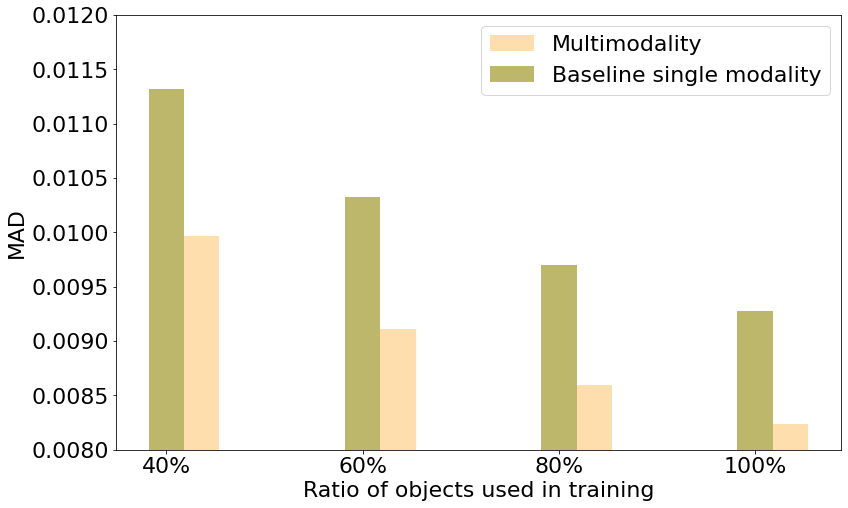}

\caption {Comparison of the multimodality impact on the MAD of the redshift estimation in the HSC 9-band dataset for four different sizes of the training set.}
\label{fig:impact_on_different_training_dataset_sizes}
\end{figure}

We examined the effect of multi-modality for various sizes of training set using the HSC 9 band dataset. Fig.~\ref{fig:impact_on_different_training_dataset_sizes} presents the results when training on  40\%, 60\%, 80\% and 100\% of the training set. The results show that the multimodality  improvement relative to the baseline remains consistent regardless of the training set size. We conclude that the effectiveness of multimodality is independent of the number of training objects.

\subsection{Multimodality impact on training}

The positive impact of multimodality can have different explanations. The most intuitive interpretation is that each parallel block that processes a subset of the input bands specializes in extracting information from  the correlations between those bands,  ultimately allowing the network to capture more relevant information than the baseline model. 
 
Alternatively,  noise may be present in the correlations between  all the bands, causing an overfit. This noise would not have a consistent relation with the redshift but the network could map it to the specific redshifts of the training sources, allowing it to optimize the training loss at the expense of extracting more general features. This would result in an under optimal performance on the validation set. Unlike the baseline, the multimodal network would avoid over-fitting this noise as the correlations between all the bands are not directly available, and so this optimization path would be  more difficult to attain.

To investigate which of these two mechanisms better explains the observed gain, we design the following experiment using the HSC 9 band dataset: we add a new modality containing all 9 images to the existing modalities of the multimodal network. If the noise present in the correlation between the bands, which is preserved in the added modality, offers the easiest optimization path and facilitates over-fitting,
we would expect the performance to degrade back to the baseline model. If, on the other hand, the benefit of multimodality arises from improved extraction of information, the additional modality should have little impact on the performance.

\begin{figure}
\centering
\includegraphics[width=\linewidth]{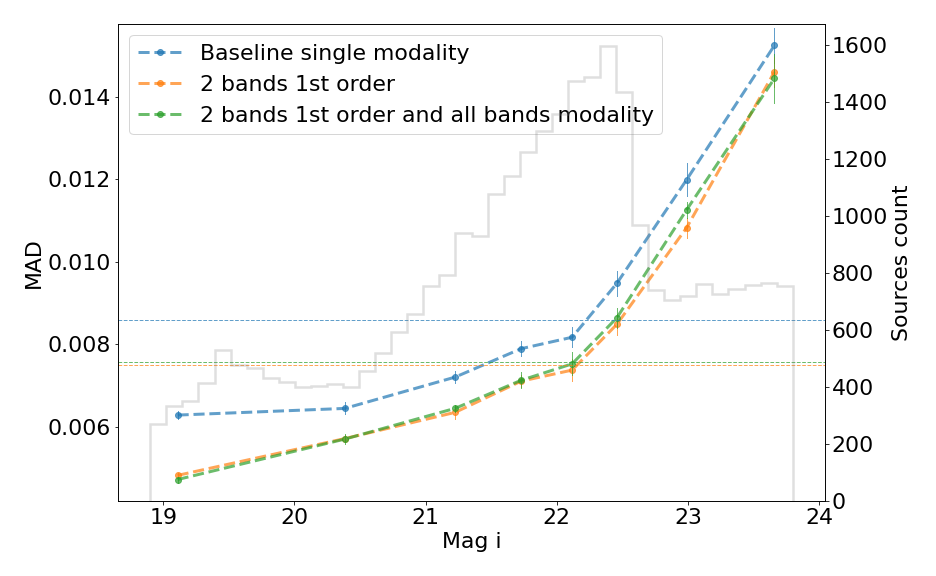}

\caption {MAD of the redshift estimation for two-band, first-order modalities with and without an additional modality containing all the bands, compared to the baseline model.
%Comparison of the MAD of the redshift estimation of three experiments. The first experiment uses a baseline approach with no multimodality, the second experiment combines two band first-order modalities with a modality that contains all the bands while the third uses two band first-order modalities. 
}
\label{fig:no_overfit_proof}
\end{figure}
The results presented in Fig. ~\ref{fig:no_overfit_proof} point to the latter option.
%model containing the additional modality with the 9 bands performs similarly to the original multimodal model. 
We conclude that the multimodality approach gains from extracting more information rather than from reducing over-fitting.

\begin{figure*}[ht!]

\centering
\includegraphics[width=\linewidth]{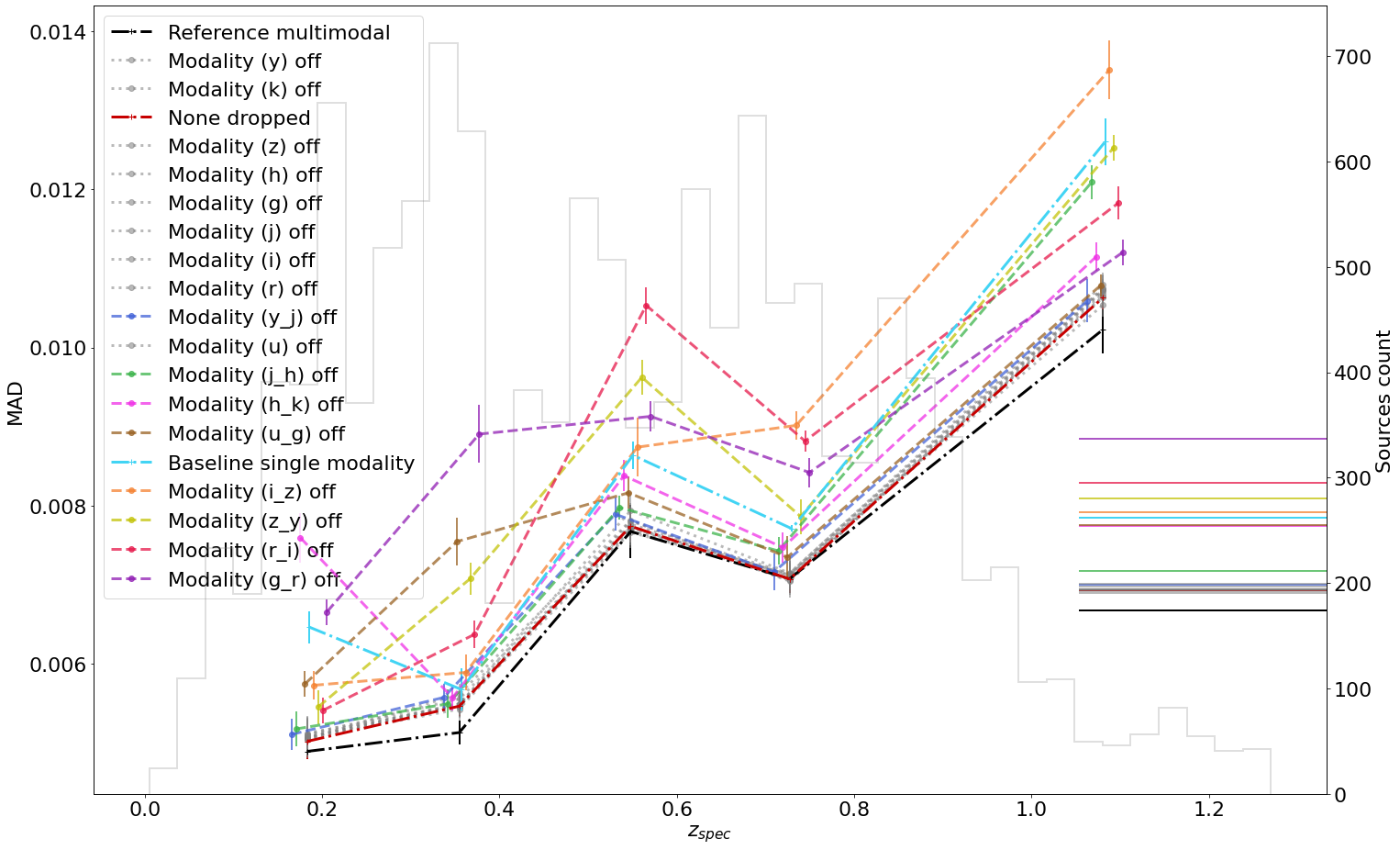}

\caption {
Evolution of the MAD as a function of redshift for the multimodal model with one modality dropped at a time. All the single-band modalities dropped are shown with grey lines as they show similar performances. We also include, for comparison,  the baseline model (single modality, cyan dot-dashed line), the reference multimodal model (black dot-dashed line) and the multimodal model trained with dropout but with no modality dropped in the test (dark red dot-dashed line). Finally, we have the different 2-band modalities dropped one at a time. Labels on the left panel are ranked according to their mean MAD. The grey histogram shows the redshift distribution of the HSC 9-band test sample. The test objects are evenly distributed between the line points which are slightly shifted on the x axis for better visual distinctiveness. The horizontal lines on the right represent the mean MAD for each experiment.}

\label{fig:modality_dropout}
\end{figure*}

\subsection{Modality dropout}
In order to study the impact of specific inter-band correlations on redshift estimation, we use a specific type of dropout technique, whereby the output of a given modality is entirely dropped, allowing us to weigh its relative importance on network performance. We aim to study the impact of the correlations between each two bands and not necessarily the bands themselves.

To do so, we train a network with two-band, first-order modalities and nine single-band modalities, to guarantee that no information is lost in the test phase when a 2-band modality is dropped. During the training phase, we randomly drop out from 0 to 5 modalities for each batch, while during the test, we consistently drop one specific modality. 

Figure~\ref{fig:modality_dropout} shows how the test MAD is affected as a function of redshift. The models are ranked according to their impact on the mean MAD value.  
We also show for comparison the baseline model, our reference multimodal model and the multimodal model trained with modality dropouts, but tested with no modality dropped. The results can be summarized as follows: 
\begin{itemize}
\item The network trained with dropouts but tested without performs similarly (marginally lower) to the reference multimodal model.  It reflects that the classical model keeps a good level of generalization. 

\item When dropping only one single-band modality, the results are also very close to the reference multimodal model whatever the band dropped. This shows that the network focuses more on the two-band modalities, as we might expect.

\item When dropping a 2-band modality, the impact is very dependant on which one is dropped.  Modalities with optical bands, $g$\_$r$ and $r$\_$i$, are overall the most important, with noticeable trends with redshift. 

The blue modality, $u$\_$g$, is critical at low redshift $z\le0.5$, while $j$\_$h$, $i$\_$z$ and $z$\_$y$ are more important at high redshift ($z\ge 1$). 

\end{itemize}
To conclude, the modality dropout test allows us to confirm that our multimodal model retains a good level of generalization and to highlight the importance of specific pairs of bands at different redshifts, as do SED fitting methods but with much better accuracy.

\section{Conclusion}

We introduce multimodality as a novel approach for redshift estimation in the framework of supervised deep learning. The input consists of galaxy images in several broad band filters, labelled with a spectroscopic redshift. Subsets of bands (modalities) are first processed separately in parallel, their respective feature maps are then combined at an appropriate stage in the network and fed to a common block. We find that this technique  enhances the extraction of color information independently of the network number of parameters and it \textit{significantly} improves the redshift precision on various datasets covering a range of characteristics (depth, sky coverage, resolution). In particular, our approach achieves new state-of-the-art results on the widely used SDSS MGS dataset.

We explored modalities of different sizes and different wavelength proximity with different stages of fusion. We conclude that the early fusion 
of modalities composed of two adjacent bands offer the best results with minimal complexity. 

Like other CNNs, our multimodal network fully exploits the information present at the pixel level but the prior parallel processing of bi-color modalities captures additional color information that improve its outcome. We find that the improvement in photometric redshift precision is statistically significant, does not depend on a specific CNN architecture, and that it increases with the number of photometric filters available. This scheme, combined with a  modality dropout test, allows us to highlight the impact of individual colors on the redshift estimation as a function of redshift.

Future work will focus on leveraging the advancements made in this study to produce redshifts for the entire HSC dataset. This will present a number of challenges, such as domain mismatch between different multi-band image acquisition conditions  and the scarcity of spectroscopically confirmed redshifts. Despite these challenges, the use of multimodality and other developped deep learning techniques have the potential to provide reliable estimates of photometric redshift, which will deliver valuable insights into the large-scale structure of the universe and the evolution of galaxies.

\begin{acknowledgements}
This work was carried out thanks to the support of the DEEPDIP ANR project (ANR-19-CE31-0023, \url{http://deepdip.net}), the Programme National Cosmologie et Galaxies (PNCG) of CNRS/INSU with INP and IN2P3, co-funded by CEA and CNES. \\
This publication makes use of Sloan Digital Sky Survey (SDSS) data. Funding for SDSS-III has been provided by the Alfred P. Sloan Foundation, the Participating Institutions, the National Science Foundation, and the U.S. Department of Energy Office of Science. The SDSS-III web site is \url{http://www.sdss3.org/}.
SDSS-III is managed by the Astrophysical Research Consortium for the Participating Institutions of the SDSS-III Collaboration including the University of Arizona, the Brazilian Participation Group, Brookhaven National Laboratory, Carnegie Mellon University, University of Florida, the French Participation Group, the German Participation Group, Harvard University, the Instituto de Astrofisica de Canarias, the Michigan State/Notre Dame/JINA Participation Group, Johns Hopkins University, Lawrence Berkeley National Laboratory, Max Planck Institute for Astrophysics, Max Planck Institute for Extraterrestrial Physics, New Mexico State University, New York University, Ohio State University, Pennsylvania State University, University of Portsmouth, Princeton University, the Spanish Participation Group, University of Tokyo, University of Utah, Vanderbilt University, University of Virginia, University of Washington, and Yale University.\\
 This publication makes use of the Canada France Hawaii Telescope Legacy Survey (CFHTLS) and CFHT Large Area U-band Deep Survey (CLAUDS), based on observations obtained with MegaPrime/MegaCam, a joint project of CFHT and CEA/IRFU, at the Canada-France-Hawaii Telescope (CFHT) which is operated by the National Research Council (NRC) of Canada, the Institut National des Science de l'Univers of the Centre National de la Recherche Scientifique (CNRS) of France, and the University of Hawaii. This work is based in part on data products produced at Terapix available at the Canadian Astronomy Data Centre as part of the Canada-France-Hawaii Telescope Legacy Survey, a collaborative project of NRC and CNRS.\\
 CLAUDS use data from the Hyper Suprime-Cam (HSC) camera. The HSC instrumentation and software were developed by the National Astronomical Observatory of Japan (NAOJ), the Kavli Institute for the Physics and Mathematics of the Universe (Kavli IPMU), the University of Tokyo, the High Energy Accelerator Research Organization (KEK), the
Academia Sinica Institute for Astronomy and Astrophysics in Taiwan (ASIAA),
and Princeton University. Funding was contributed by the FIRST program from
Japanese Cabinet Office, the Ministry of Education, Culture, Sports, Science
and Technology (MEXT), the Japan Society for the Promotion of Science
(JSPS), Japan Science and Technology Agency (JST), the Toray Science Foundation, NAOJ, Kavli IPMU, KEK, ASIAA, and Princeton University.
CLAUDS is a collaboration between astronomers from Canada, France, and China described in Sawicki et al. (2019) and uses data products from CALET and the Canadian Astronomy Data Centre (CADC) and was processed using resources from Compute Canada and Canadian Advanced Network For Astrophysical Research (CANFAR) and the CANDIDE cluster at
IAP maintained by Stephane Rouberol.
\end{acknowledgements} 

\bibliographystyle{aa}
\bibliography{references_Reda}

\newcommand{\noop}[1]{}
\begin{thebibliography}{71}
\expandafter\ifx\csname natexlab\endcsname\relax\def\natexlab#1{#1}\fi

\bibitem[{{Ahumada} {et~al.}(2020){Ahumada}, {Prieto}, {Almeida}, {Anders}, {Anderson}, {Andrews}, {Anguiano}, {Arcodia}, {Armengaud}, {Aubert}, \& et~al.}]{SDSS_DR16}
{Ahumada}, R., {Prieto}, C.~A., {Almeida}, A., {et~al.} 2020, \apjs, 249, 3

\bibitem[{{Aihara} {et~al.}(2019){Aihara}, {AlSayyad}, {Ando}, {Armstrong}, {Bosch}, {Egami}, {Furusawa}, {Furusawa}, {Goulding}, {Harikane}, {Hikage}, {Ho}, {Hsieh}, {Huang}, {Ikeda}, {Imanishi}, {Ito}, {Iwata}, {Jaelani}, {Kakuma}, {Kawana}, {Kikuta}, {Kobayashi}, {Koike}, {Komiyama}, {Li}, {Liang}, {Lin}, {Luo}, {Lupton}, {Lust}, {MacArthur}, {Matsuoka}, {Mineo}, {Miyatake}, {Miyazaki}, {More}, {Murata}, {Namiki}, {Nishizawa}, {Oguri}, {Okabe}, {Okamoto}, {Okura}, {Ono}, {Onodera}, {Onoue}, {Osato}, {Ouchi}, {Shibuya}, {Strauss}, {Sugiyama}, {Suto}, {Takada}, {Takagi}, {Takata}, {Takita}, {Tanaka}, {Terai}, {Toba}, {Uchiyama}, {Utsumi}, {Wang}, {Wang}, \& {Yamada}}]{Aihara2019}
{Aihara}, H., {AlSayyad}, Y., {Ando}, M., {et~al.} 2019, \pasj, 106

\bibitem[{{Alam} {et~al.}(2015){Alam}, {Albareti}, {Allende Prieto}, {Anders}, {Anderson}, {Anderton}, {Andrews}, {Armengaud}, {Aubourg}, {Bailey}, \& et~al.}]{Alam2015}
{Alam}, S., {Albareti}, F.~D., {Allende Prieto}, C., {et~al.} 2015, \apjs, 219, 12

\bibitem[{{Arnouts} {et~al.}(1999){Arnouts}, {Cristiani}, {Moscardini}, {Matarrese}, {Lucchin}, {Fontana}, \& {Giallongo}}]{Arnouts1999}
{Arnouts}, S., {Cristiani}, S., {Moscardini}, L., {et~al.} 1999, \mnras, 310, 540

\bibitem[{{Baldry} {et~al.}(2018){Baldry}, {Liske}, {Brown}, {Robotham}, {Driver}, {Dunne}, {Alpaslan}, {Brough}, {Cluver}, {Eardley}, {Farrow}, {Heymans}, {Hildebrandt}, {Hopkins}, {Kelvin}, {Loveday}, {Moffett}, {Norberg}, {Owers}, {Taylor}, {Wright}, {Bamford}, {Bland-Hawthorn}, {Bourne}, {Bremer}, {Colless}, {Conselice}, {Croom}, {Davies}, {Foster}, {Grootes}, {Holwerda}, {Jones}, {Kafle}, {Kuijken}, {Lara-Lopez}, {L{\'o}pez-S{\'a}nchez}, {Meyer}, {Phillipps}, {Sutherland}, {van Kampen}, \& {Wilkins}}]{Baldry2018}
{Baldry}, I.~K., {Liske}, J., {Brown}, M.~J.~I., {et~al.} 2018, \mnras, 474, 3875

\bibitem[{{Beck} {et~al.}(2016){Beck}, {Dobos}, {Budav{\'a}ri}, {Szalay}, \& {Csabai}}]{Beck2016}
{Beck}, R., {Dobos}, L., {Budav{\'a}ri}, T., {Szalay}, A.~S., \& {Csabai}, I. 2016, \mnras, 460, 1371

\bibitem[{{Ben{\'\i}tez}(2000)}]{Benitez2000}
{Ben{\'\i}tez}, N. 2000, \apj, 536, 571

\bibitem[{Berg-Kirkpatrick {et~al.}(2012)Berg-Kirkpatrick, Burkett, \& Klein}]{berg-kirkpatrick-etal-2012-empirical}
Berg-Kirkpatrick, T., Burkett, D., \& Klein, D. 2012, in Proceedings of the 2012 Joint Conference on Empirical Methods in Natural Language Processing and Computational Natural Language Learning (Jeju Island, Korea: Association for Computational Linguistics), 995--1005

\bibitem[{Bertin(2006)}]{Bertin2006}
Bertin, E. 2006, Astronomical Data Analysis Software and Systems XV ASP Conference Series, 351, 112

\bibitem[{{Bertin} \& {Arnouts}(1996)}]{Bertin1996}
{Bertin}, E. \& {Arnouts}, S. 1996, \aaps, 117, 393

\bibitem[{{Bertin} {et~al.}(2002){Bertin}, {Mellier}, {Radovich}, {Missonnier}, {Didelon}, \& {Morin}}]{2002ASPC..281..228B}
{Bertin}, E., {Mellier}, Y., {Radovich}, M., {et~al.} 2002, in Astronomical Society of the Pacific Conference Series, Vol. 281, Astronomical Data Analysis Software and Systems XI, ed. D.~A. {Bohlender}, D.~{Durand}, \& T.~H. {Handley}, 228

\bibitem[{{Boulade} {et~al.}(2000){Boulade}, {Charlot}, {Abbon}, {Aune}, {Borgeaud}, {Carton}, {Carty}, {Desforge}, {Eppele}, {Gallais}, {Gosset}, {Granelli}, {Gros}, {de Kat}, {Loiseau}, {Mellier}, {Ritou}, {Rousse}, {Starzynski}, {Vignal}, \& {Vigroux}}]{Boulade2000}
{Boulade}, O., {Charlot}, X., {Abbon}, P., {et~al.} 2000, in Society of Photo-Optical Instrumentation Engineers (SPIE) Conference Series, Vol. 4008, Society of Photo-Optical Instrumentation Engineers (SPIE) Conference Series, ed. M.~{Iye} \& A.~F. {Moorwood}, 657--668

\bibitem[{{Bradshaw} {et~al.}(2013){Bradshaw}, {Almaini}, {Hartley}, {Smith}, {Conselice}, {Dunlop}, {Simpson}, {Chuter}, {Cirasuolo}, {Foucaud}, {McLure}, {Mortlock}, \& {Pearce}}]{Bradshaw2013}
{Bradshaw}, E.~J., {Almaini}, O., {Hartley}, W.~G., {et~al.} 2013, \mnras, 433, 194

\bibitem[{{Brammer} {et~al.}(2008){Brammer}, {van Dokkum}, \& {Coppi}}]{Brammer2008}
{Brammer}, G.~B., {van Dokkum}, P.~G., \& {Coppi}, P. 2008, \apj, 686, 1503

\bibitem[{Carliles {et~al.}(2010)Carliles, Budav{\'a}ri, Heinis, Priebe, \& Szalay}]{carliles2010random}
Carliles, S., Budav{\'a}ri, T., Heinis, S., Priebe, C., \& Szalay, A.~S. 2010, The Astrophysical Journal, 712, 511

\bibitem[{Chen {et~al.}(2017)Chen, Li, Ghamisi, Jia, \& Gu}]{chen2017deep}
Chen, Y., Li, C., Ghamisi, P., Jia, X., \& Gu, Y. 2017, IEEE Geoscience and Remote Sensing Letters, 14, 1253

\bibitem[{{Coil} {et~al.}(2011){Coil}, {Blanton}, {Burles}, {Cool}, {Eisenstein}, {Moustakas}, {Wong}, {Zhu}, {Aird}, {Bernstein}, {Bolton}, \& {Hogg}}]{Coil2011}
{Coil}, A.~L., {Blanton}, M.~R., {Burles}, S.~M., {et~al.} 2011, \apj, 741, 8

\bibitem[{{Collister} \& {Lahav}(2004)}]{Collister2004}
{Collister}, A.~A. \& {Lahav}, O. 2004, \pasp, 116, 345

\bibitem[{{Cool} {et~al.}(2013){Cool}, {Moustakas}, {Blanton}, {Burles}, {Coil}, {Eisenstein}, {Wong}, {Zhu}, {Aird}, {Bernstein}, {Bolton}, {Hogg}, \& {Mendez}}]{Cool2013}
{Cool}, R.~J., {Moustakas}, J., {Blanton}, M.~R., {et~al.} 2013, \apj, 767, 118

\bibitem[{{Csabai} {et~al.}(2007){Csabai}, {Dobos}, {Trencs{\'e}ni}, {Herczegh}, {J{\'o}zsa}, {Purger}, {Budav{\'a}ri}, \& {Szalay}}]{Csabai2007}
{Csabai}, I., {Dobos}, L., {Trencs{\'e}ni}, M., {et~al.} 2007, Astronomische Nachrichten, 328, 852

\bibitem[{{Dalton} {et~al.}(2006){Dalton}, {Caldwell}, {Ward}, {Whalley}, {Woodhouse}, {Edeson}, {Clark}, {Beard}, {Gallie}, {Todd}, {Strachan}, {Bezawada}, {Sutherland}, \& {Emerson}}]{Dalton2006}
{Dalton}, G.~B., {Caldwell}, M., {Ward}, A.~K., {et~al.} 2006, in Society of Photo-Optical Instrumentation Engineers (SPIE) Conference Series, Vol. 6269, Society of Photo-Optical Instrumentation Engineers (SPIE) Conference Series, ed. I.~S. {McLean} \& M.~{Iye}, 62690X

\bibitem[{{de Jong} {et~al.}(2013){de Jong}, {Kuijken}, {Applegate}, {Begeman}, {Belikov}, {Blake}, {Bout}, {Boxhoorn}, {Buddelmeijer}, {Buddendiek}, {Cacciato}, {Capaccioli}, {Choi}, {Cordes}, {Covone}, {Dall'Ora}, {Edge}, {Erben}, {Franse}, {Getman}, {Grado}, {Harnois-Deraps}, {Helmich}, {Herbonnet}, {Heymans}, {Hildebrand t}, {Hoekstra}, {Huang}, {Irisarri}, {Joachimi}, {K{\"o}hlinger}, {Kitching}, {La Barbera}, {Lacerda}, {McFarland}, {Miller}, {Nakajima}, {Napolitano}, {Paolillo}, {Peacock}, {Pila-Diez}, {Puddu}, {Radovich}, {Rifatto}, {Schneider}, {Schrabback}, {Sifon}, {Sikkema}, {Simon}, {Sutherland}, {Tudorica}, {Valentijn}, {van der Burg}, {van Uitert}, {van Waerbeke}, {Veland er}, {Verdoes Kleijn}, {Viola}, \& {Vriend}}]{deJong2013}
{de Jong}, J.~T.~A., {Kuijken}, K., {Applegate}, D., {et~al.} 2013, The Messenger, 154, 44

\bibitem[{{Desprez} {et~al.}(2023){Desprez}, {Picouet}, {Moutard}, {Arnouts}, {Sawicki}, {Coupon}, {Gwyn}, {Chen}, {Huang}, {Golob}, {Furusawa}, {Ikeda}, {Paltani}, {Cheng}, {Hartley}, {Hsieh}, {Ilbert}, {Kauffmann}, {McCracken}, {Shuntov}, {Tanaka}, {Toft}, {Tresse}, \& {Weaver}}]{Desprez2023}
{Desprez}, G., {Picouet}, V., {Moutard}, T., {et~al.} 2023, \aap, 670, A82

\bibitem[{{Dey} {et~al.}(2021){Dey}, {Andrews}, {Newman}, {Mao}, {Rau}, \& {Zhou}}]{Dey2021}
{Dey}, B., {Andrews}, B.~H., {Newman}, J.~A., {et~al.} 2021, arXiv e-prints, arXiv:2112.03939

\bibitem[{{Drinkwater} {et~al.}(2018){Drinkwater}, {Byrne}, {Blake}, {Glazebrook}, {Brough}, {Colless}, {Couch}, {Croton}, {Croom}, {Davis}, {Forster}, {Gilbank}, {Hinton}, {Jelliffe}, {Jurek}, {Li}, {Martin}, {Pimbblet}, {Poole}, {Pracy}, {Sharp}, {Smillie}, {Spolaor}, {Wisnioski}, {Woods}, {Wyder}, \& {Yee}}]{Drinkwater2018}
{Drinkwater}, M.~J., {Byrne}, Z.~J., {Blake}, C., {et~al.} 2018, \mnras, 474, 4151

\bibitem[{Efron \& Tibshirani(1994)}]{efron1994introduction}
Efron, B. \& Tibshirani, R.~J. 1994, An introduction to the bootstrap (CRC press)

\bibitem[{{Emerson} {et~al.}(2004){Emerson}, {Sutherland}, {McPherson}, {Craig}, {Dalton}, \& {Ward}}]{Emerson2004}
{Emerson}, J.~P., {Sutherland}, W.~J., {McPherson}, A.~M., {et~al.} 2004, The Messenger, 117, 27

\bibitem[{{Garilli} {et~al.}(2021){Garilli}, {McLure}, {Pentericci}, {Franzetti}, {Gargiulo}, {Carnall}, {Cucciati}, {Iovino}, {Amorin}, {Bolzonella}, {Bongiorno}, {Castellano}, {Cimatti}, {Cirasuolo}, {Cullen}, {Dunlop}, {Elbaz}, {Finkelstein}, {Fontana}, {Fontanot}, {Fumana}, {Guaita}, {Hartley}, {Jarvis}, {Juneau}, {Maccagni}, {McLeod}, {Nandra}, {Pompei}, {Pozzetti}, {Scodeggio}, {Talia}, {Calabr{\`o}}, {Cresci}, {Fynbo}, {Hathi}, {Hibon}, {Koekemoer}, {Magliocchetti}, {Salvato}, {Vietri}, {Zamorani}, {Almaini}, {Balestra}, {Bardelli}, {Begley}, {Brammer}, {Bell}, {Bowler}, {Brusa}, {Buitrago}, {Caputi}, {Cassata}, {Charlot}, {Citro}, {Cristiani}, {Curtis-Lake}, {Dickinson}, {Fazio}, {Ferguson}, {Fiore}, {Franco}, {Georgakakis}, {Giavalisco}, {Grazian}, {Hamadouche}, {Jung}, {Kim}, {Khusanova}, {Le F{\`e}vre}, {Longhetti}, {Lotz}, {Mannucci}, {Maltby}, {Matsuoka}, {Mendez-Hernandez}, {Mendez-Abreu}, {Mignoli}, {Moresco}, {Nonino}, {Pannella}, {Papovich}, {Popesso}, {Roberts-Borsani}, {Rosario},
  {Saldana-Lopez}, {Santini}, {Saxena}, {Schaerer}, {Schreiber}, {Stark}, {Tasca}, {Thomas}, {Vanzella}, {Wild}, {Williams}, \& {Zucca}}]{Garilli2021}
{Garilli}, B., {McLure}, R., {Pentericci}, L., {et~al.} 2021, \aap, 647, A150

\bibitem[{Gneiting \& Raftery(2007)}]{gneiting2007strictly}
Gneiting, T. \& Raftery, A.~E. 2007, Journal of the American statistical Association, 102, 359

\bibitem[{Goodfellow {et~al.}(2016)Goodfellow, Bengio, \& Courville}]{goodfellow2016deep}
Goodfellow, I., Bengio, Y., \& Courville, A. 2016, Deep learning (MIT press)

\bibitem[{{Hayat} {et~al.}(2021){Hayat}, {Stein}, {Harrington}, {Luki{\'c}}, \& {Mustafa}}]{Hayat2021}
{Hayat}, M.~A., {Stein}, G., {Harrington}, P., {Luki{\'c}}, Z., \& {Mustafa}, M. 2021, \apjl, 911, L33

\bibitem[{He {et~al.}(2015)He, Zhang, Ren, \& Sun}]{he2015delving}
He, K., Zhang, X., Ren, S., \& Sun, J. 2015, in Proceedings of the IEEE international conference on computer vision, 1026--1034

\bibitem[{Hong {et~al.}(2020)Hong, Gao, Yokoya, Yao, Chanussot, Du, \& Zhang}]{hong2020more}
Hong, D., Gao, L., Yokoya, N., {et~al.} 2020, IEEE Transactions on Geoscience and Remote Sensing, 59, 4340

\bibitem[{Hou {et~al.}(2018)Hou, Wang, Lai, Tsao, Chang, \& Wang}]{hou2018audio}
Hou, J.-C., Wang, S.-S., Lai, Y.-H., {et~al.} 2018, IEEE Transactions on Emerging Topics in Computational Intelligence, 2, 117

\bibitem[{{Hudelot} {et~al.}(2012){Hudelot}, {Cuillandre}, {Withington}, {Goranova}, {McCracken}, {Magnard}, {Mellier}, {Regnault}, {Betoule}, {Aussel}, {Kavelaars}, {Fernique}, {Bonnarel}, {Ochsenbein}, \& {Ilbert}}]{Hudelot2012}
{Hudelot}, P., {Cuillandre}, J.~C., {Withington}, K., {et~al.} 2012, VizieR Online Data Catalog, II/317

\bibitem[{{Ilbert} {et~al.}(2006){Ilbert}, {Arnouts}, {McCracken}, {Bolzonella}, {Bertin}, {Le F{\`e}vre}, {Mellier}, {Zamorani}, {Pell{\`o}}, {Iovino}, {Tresse}, {Le Brun}, {Bottini}, {Garilli}, {Maccagni}, {Picat}, {Scaramella}, {Scodeggio}, {Vettolani}, {Zanichelli}, {Adami}, {Bardelli}, {Cappi}, {Charlot}, {Ciliegi}, {Contini}, {Cucciati}, {Foucaud}, {Franzetti}, {Gavignaud}, {Guzzo}, {Marano}, {Marinoni}, {Mazure}, {Meneux}, {Merighi}, {Paltani}, {Pollo}, {Pozzetti}, {Radovich}, {Zucca}, {Bondi}, {Bongiorno}, {Busarello}, {de La Torre}, {Gregorini}, {Lamareille}, {Mathez}, {Merluzzi}, {Ripepi}, {Rizzo}, \& {Vergani}}]{Ilbert2006}
{Ilbert}, O., {Arnouts}, S., {McCracken}, H.~J., {et~al.} 2006, \aap, 457, 841

\bibitem[{{Ivezi{\'c}} {et~al.}(2019){Ivezi{\'c}}, {Kahn}, {Tyson}, {Abel}, {Acosta}, {Allsman}, {Alonso}, {AlSayyad}, {Anderson}, {Andrew}, {Angel}, {Angeli}, {Ansari}, {Antilogus}, {Araujo}, {Armstrong}, {Arndt}, {Astier}, {Aubourg}, {Auza}, {Axelrod}, {Bard}, {Barr}, {Barrau}, {Bartlett}, {Bauer}, {Bauman}, {Baumont}, {Bechtol}, {Bechtol}, {Becker}, {Becla}, {Beldica}, {Bellavia}, {Bianco}, {Biswas}, {Blanc}, {Blazek}, {Bland ford}, {Bloom}, {Bogart}, {Bond}, {Booth}, {Borgland}, {Borne}, {Bosch}, {Boutigny}, {Brackett}, {Bradshaw}, {Brand t}, {Brown}, {Bullock}, {Burchat}, {Burke}, {Cagnoli}, {Calabrese}, {Callahan}, {Callen}, {Carlin}, {Carlson}, {Chand rasekharan}, {Charles-Emerson}, {Chesley}, {Cheu}, {Chiang}, {Chiang}, {Chirino}, {Chow}, {Ciardi}, {Claver}, {Cohen-Tanugi}, {Cockrum}, {Coles}, {Connolly}, {Cook}, {Cooray}, {Covey}, {Cribbs}, {Cui}, {Cutri}, {Daly}, {Daniel}, {Daruich}, {Daubard}, {Daues}, {Dawson}, {Delgado}, {Dellapenna}, {de Peyster}, {de Val-Borro}, {Digel}, {Doherty}, {Dubois},
  {Dubois-Felsmann}, {Durech}, {Economou}, {Eifler}, {Eracleous}, {Emmons}, {Fausti Neto}, {Ferguson}, {Figueroa}, {Fisher-Levine}, {Focke}, {Foss}, {Frank}, {Freemon}, {Gangler}, {Gawiser}, {Geary}, {Gee}, {Geha}, {Gessner}, {Gibson}, {Gilmore}, {Glanzman}, {Glick}, {Goldina}, {Goldstein}, {Goodenow}, {Graham}, {Gressler}, {Gris}, {Guy}, {Guyonnet}, {Haller}, {Harris}, {Hascall}, {Haupt}, {Hernand ez}, {Herrmann}, {Hileman}, {Hoblitt}, {Hodgson}, {Hogan}, {Howard}, {Huang}, {Huffer}, {Ingraham}, {Innes}, {Jacoby}, {Jain}, {Jammes}, {Jee}, {Jenness}, {Jernigan}, {Jevremovi{\'c}}, {Johns}, {Johnson}, {Johnson}, {Jones}, {Juramy-Gilles}, {Juri{\'c}}, {Kalirai}, {Kallivayalil}, {Kalmbach}, {Kantor}, {Karst}, {Kasliwal}, {Kelly}, {Kessler}, {Kinnison}, {Kirkby}, {Knox}, {Kotov}, {Krabbendam}, {Krughoff}, {Kub{\'a}nek}, {Kuczewski}, {Kulkarni}, {Ku}, {Kurita}, {Lage}, {Lambert}, {Lange}, {Langton}, {Le Guillou}, {Levine}, {Liang}, {Lim}, {Lintott}, {Long}, {Lopez}, {Lotz}, {Lupton}, {Lust}, {MacArthur}, {Mahabal},
  {Mand elbaum}, {Markiewicz}, {Marsh}, {Marshall}, {Marshall}, {May}, {McKercher}, {McQueen}, {Meyers}, {Migliore}, {Miller}, {Mills}, {Miraval}, {Moeyens}, {Moolekamp}, {Monet}, {Moniez}, {Monkewitz}, {Montgomery}, {Morrison}, {Mueller}, {Muller}, {Mu{\~n}oz Arancibia}, {Neill}, {Newbry}, {Nief}, {Nomerotski}, {Nordby}, {O'Connor}, {Oliver}, {Olivier}, {Olsen}, {O'Mullane}, {Ortiz}, {Osier}, {Owen}, {Pain}, {Palecek}, {Parejko}, {Parsons}, {Pease}, {Peterson}, {Peterson}, {Petravick}, {Libby Petrick}, {Petry}, {Pierfederici}, {Pietrowicz}, {Pike}, {Pinto}, {Plante}, {Plate}, {Plutchak}, {Price}, {Prouza}, {Radeka}, {Rajagopal}, {Rasmussen}, {Regnault}, {Reil}, {Reiss}, {Reuter}, {Ridgway}, {Riot}, {Ritz}, {Robinson}, {Roby}, {Roodman}, {Rosing}, {Roucelle}, {Rumore}, {Russo}, {Saha}, {Sassolas}, {Schalk}, {Schellart}, {Schindler}, {Schmidt}, {Schneider}, {Schneider}, {Schoening}, {Schumacher}, {Schwamb}, {Sebag}, {Selvy}, {Sembroski}, {Seppala}, {Serio}, {Serrano}, {Shaw}, {Shipsey}, {Sick}, {Silvestri},
  {Slater}, {Smith}, {Smith}, {Sobhani}, {Soldahl}, {Storrie-Lombardi}, {Stover}, {Strauss}, {Street}, {Stubbs}, {Sullivan}, {Sweeney}, {Swinbank}, {Szalay}, {Takacs}, {Tether}, {Thaler}, {Thayer}, {Thomas}, {Thornton}, {Thukral}, {Tice}, {Trilling}, {Turri}, {Van Berg}, {Vanden Berk}, {Vetter}, {Virieux}, {Vucina}, {Wahl}, {Walkowicz}, {Walsh}, {Walter}, {Wang}, {Wang}, {Warner}, {Wiecha}, {Willman}, {Winters}, {Wittman}, {Wolff}, {Wood-Vasey}, {Wu}, {Xin}, {Yoachim}, \& {Zhan}}]{Ivezic2019}
{Ivezi{\'c}}, {\v{Z}}., {Kahn}, S.~M., {Tyson}, J.~A., {et~al.} 2019, \apj, 873, 111

\bibitem[{{Jarvis} {et~al.}(2013){Jarvis}, {Bonfield}, {Bruce}, {Geach}, {McAlpine}, {McLure}, {Gonz{\'a}lez-Solares}, {Irwin}, {Lewis}, {Yoldas}, {Andreon}, {Cross}, {Emerson}, {Dalton}, {Dunlop}, {Hodgkin}, {Le}, {Karouzos}, {Meisenheimer}, {Oliver}, {Rawlings}, {Simpson}, {Smail}, {Smith}, {Sullivan}, {Sutherland}, {White}, \& {Zwart}}]{Jarvis2013}
{Jarvis}, M.~J., {Bonfield}, D.~G., {Bruce}, V.~A., {et~al.} 2013, \mnras, 428, 1281

\bibitem[{Kingma \& Ba(2014)}]{kingma2014adam}
Kingma, D.~P. \& Ba, J. 2014, arXiv preprint arXiv:1412.6980

\bibitem[{Koehn(2004)}]{koehn-2004-statistical}
Koehn, P. 2004, in Proceedings of the 2004 Conference on Empirical Methods in Natural Language Processing (Barcelona, Spain: Association for Computational Linguistics), 388--395

\bibitem[{Krizhevsky {et~al.}(2017)Krizhevsky, Sutskever, \& Hinton}]{krizhevsky2017imagenet}
Krizhevsky, A., Sutskever, I., \& Hinton, G.~E. 2017, Communications of the ACM, 60, 84

\bibitem[{{Laureijs} {et~al.}(2011){Laureijs}, {Amiaux}, {Arduini}, {Augu{\`e}res}, {Brinchmann}, {Cole}, {Cropper}, {Dabin}, {Duvet}, {Ealet}, {Garilli}, {Gondoin}, {Guzzo}, {Hoar}, {Hoekstra}, {Holmes}, {Kitching}, {Maciaszek}, {Mellier}, {Pasian}, {Percival}, {Rhodes}, {Saavedra Criado}, {Sauvage}, {Scaramella}, {Valenziano}, {Warren}, {Bender}, {Castander}, {Cimatti}, {Le F{\`e}vre}, {Kurki-Suonio}, {Levi}, {Lilje}, {Meylan}, {Nichol}, {Pedersen}, {Popa}, {Rebolo Lopez}, {Rix}, {Rottgering}, {Zeilinger}, {Grupp}, {Hudelot}, {Massey}, {Meneghetti}, {Miller}, {Paltani}, {Paulin-Henriksson}, {Pires}, {Saxton}, {Schrabback}, {Seidel}, {Walsh}, {Aghanim}, {Amendola}, {Bartlett}, {Baccigalupi}, {Beaulieu}, {Benabed}, {Cuby}, {Elbaz}, {Fosalba}, {Gavazzi}, {Helmi}, {Hook}, {Irwin}, {Kneib}, {Kunz}, {Mannucci}, {Moscardini}, {Tao}, {Teyssier}, {Weller}, {Zamorani}, {Zapatero Osorio}, {Boulade}, {Foumond}, {Di Giorgio}, {Guttridge}, {James}, {Kemp}, {Martignac}, {Spencer}, {Walton}, {Bl{\"u}mchen}, {Bonoli},
  {Bortoletto}, {Cerna}, {Corcione}, {Fabron}, {Jahnke}, {Ligori}, {Madrid}, {Martin}, {Morgante}, {Pamplona}, {Prieto}, {Riva}, {Toledo}, {Trifoglio}, {Zerbi}, {Abdalla}, {Douspis}, {Grenet}, {Borgani}, {Bouwens}, {Courbin}, {Delouis}, {Dubath}, {Fontana}, {Frailis}, {Grazian}, {Koppenh{\"o}fer}, {Mansutti}, {Melchior}, {Mignoli}, {Mohr}, {Neissner}, {Noddle}, {Poncet}, {Scodeggio}, {Serrano}, {Shane}, {Starck}, {Surace}, {Taylor}, {Verdoes-Kleijn}, {Vuerli}, {Williams}, {Zacchei}, {Altieri}, {Escudero Sanz}, {Kohley}, {Oosterbroek}, {Astier}, {Bacon}, {Bardelli}, {Baugh}, {Bellagamba}, {Benoist}, {Bianchi}, {Biviano}, {Branchini}, {Carbone}, {Cardone}, {Clements}, {Colombi}, {Conselice}, {Cresci}, {Deacon}, {Dunlop}, {Fedeli}, {Fontanot}, {Franzetti}, {Giocoli}, {Garcia-Bellido}, {Gow}, {Heavens}, {Hewett}, {Heymans}, {Holland}, {Huang}, {Ilbert}, {Joachimi}, {Jennins}, {Kerins}, {Kiessling}, {Kirk}, {Kotak}, {Krause}, {Lahav}, {van Leeuwen}, {Lesgourgues}, {Lombardi}, {Magliocchetti}, {Maguire},
  {Majerotto}, {Maoli}, {Marulli}, {Maurogordato}, {McCracken}, {McLure}, {Melchiorri}, {Merson}, {Moresco}, {Nonino}, {Norberg}, {Peacock}, {Pello}, {Penny}, {Pettorino}, {Di Porto}, {Pozzetti}, {Quercellini}, {Radovich}, {Rassat}, {Roche}, {Ronayette}, {Rossetti}, {Sartoris}, {Schneider}, {Semboloni}, {Serjeant}, {Simpson}, {Skordis}, {Smadja}, {Smartt}, {Spano}, {Spiro}, {Sullivan}, {Tilquin}, {Trotta}, {Verde}, {Wang}, {Williger}, {Zhao}, {Zoubian}, \& {Zucca}}]{Laureijs2011}
{Laureijs}, R., {Amiaux}, J., {Arduini}, S., {et~al.} 2011, arXiv e-prints, arXiv:1110.3193

\bibitem[{{Le F{\`e}vre} {et~al.}(2013){Le F{\`e}vre}, {Cassata}, {Cucciati}, {Garilli}, {Ilbert}, {Le Brun}, {Maccagni}, {Moreau}, {Scodeggio}, {Tresse}, {Zamorani}, {Adami}, {Arnouts}, {Bardelli}, {Bolzonella}, {Bondi}, {Bongiorno}, {Bottini}, {Cappi}, {Charlot}, {Ciliegi}, {Contini}, {de la Torre}, {Foucaud}, {Franzetti}, {Gavignaud}, {Guzzo}, {Iovino}, {Lemaux}, {L{\'o}pez-Sanjuan}, {McCracken}, {Marano}, {Marinoni}, {Mazure}, {Mellier}, {Merighi}, {Merluzzi}, {Paltani}, {Pell{\`o}}, {Pollo}, {Pozzetti}, {Scaramella}, {Tasca}, {Vergani}, {Vettolani}, {Zanichelli}, \& {Zucca}}]{LeFevre2013}
{Le F{\`e}vre}, O., {Cassata}, P., {Cucciati}, O., {et~al.} 2013, \aap, 559, A14

\bibitem[{{Le F{\`e}vre} {et~al.}(2015){Le F{\`e}vre}, {Tasca}, {Cassata}, {Garilli}, {Le Brun}, {Maccagni}, {Pentericci}, {Thomas}, {Vanzella}, {Zamorani}, {Zucca}, {Amorin}, {Bardelli}, {Capak}, {Cassar{\`a}}, {Castellano}, {Cimatti}, {Cuby}, {Cucciati}, {de la Torre}, {Durkalec}, {Fontana}, {Giavalisco}, {Grazian}, {Hathi}, {Ilbert}, {Lemaux}, {Moreau}, {Paltani}, {Ribeiro}, {Salvato}, {Schaerer}, {Scodeggio}, {Sommariva}, {Talia}, {Taniguchi}, {Tresse}, {Vergani}, {Wang}, {Charlot}, {Contini}, {Fotopoulou}, {L{\'o}pez-Sanjuan}, {Mellier}, \& {Scoville}}]{LeFevre2015}
{Le F{\`e}vre}, O., {Tasca}, L.~A.~M., {Cassata}, P., {et~al.} 2015, \aap, 576, A79

\bibitem[{LeCun {et~al.}(2015)LeCun, Bengio, \& Hinton}]{lecun2015deep}
LeCun, Y., Bengio, Y., \& Hinton, G. 2015, nature, 521, 436

\bibitem[{{Lee} {et~al.}(2018){Lee}, {Krolewski}, {White}, {Schlegel}, {Nugent}, {Hennawi}, {M{\"u}ller}, {Pan}, {Prochaska}, {Font-Ribera}, {Suzuki}, {Glazebrook}, {Kacprzak}, {Kartaltepe}, {Koekemoer}, {Le F{\`e}vre}, {Lemaux}, {Maier}, {Nanayakkara}, {Rich}, {Sanders}, {Salvato}, {Tasca}, \& {Tran}}]{KGLee2018}
{Lee}, K.-G., {Krolewski}, A., {White}, M., {et~al.} 2018, \apjs, 237, 31

\bibitem[{Lilly {et~al.}(2007)Lilly, Le~F{\`e}vre, Renzini, Zamorani, Scodeggio, Contini, Carollo, Hasinger, Kneib, Iovino, {et~al.}}]{Lilly2007}
Lilly, S.~J., Le~F{\`e}vre, O., Renzini, A., {et~al.} 2007, The Astrophysical Journal Supplement Series, 172, 70

\bibitem[{Ma {et~al.}(2015)Ma, Lu, Shang, \& Li}]{ma2015multimodal}
Ma, L., Lu, Z., Shang, L., \& Li, H. 2015, in Proceedings of the IEEE international conference on computer vision, 2623--2631

\bibitem[{{Masters} {et~al.}(2019){Masters}, {Stern}, {Cohen}, {Capak}, {Stanford}, {Hernitschek}, {Galametz}, {Davidzon}, {Rhodes}, {Sand ers}, {Mobasher}, {Castander}, {Pruett}, \& {Fotopoulou}}]{Masters2019}
{Masters}, D.~C., {Stern}, D.~K., {Cohen}, J.~G., {et~al.} 2019, \apj, 877, 81

\bibitem[{{McCracken} {et~al.}(2012){McCracken}, {Milvang-Jensen}, {Dunlop}, {Franx}, {Fynbo}, {Le F{\`e}vre}, {Holt}, {Caputi}, {Goranova}, {Buitrago}, {Emerson}, {Freudling}, {Hudelot}, {L{\'o}pez-Sanjuan}, {Magnard}, {Mellier}, {M{\o}ller}, {Nilsson}, {Sutherland}, {Tasca}, \& {Zabl}}]{McCracken2012}
{McCracken}, H.~J., {Milvang-Jensen}, B., {Dunlop}, J., {et~al.} 2012, \aap, 544, A156

\bibitem[{{McLure} {et~al.}(2013){McLure}, {Pearce}, {Dunlop}, {Cirasuolo}, {Curtis-Lake}, {Bruce}, {Caputi}, {Almaini}, {Bonfield}, {Bradshaw}, {Buitrago}, {Chuter}, {Foucaud}, {Hartley}, \& {Jarvis}}]{McLure2013}
{McLure}, R.~J., {Pearce}, H.~J., {Dunlop}, J.~S., {et~al.} 2013, \mnras, 428, 1088

\bibitem[{{Miyazaki} {et~al.}(2018){Miyazaki}, {Komiyama}, {Kawanomoto}, {Doi}, {Furusawa}, {Hamana}, {Hayashi}, {Ikeda}, {Kamata}, {Karoji}, {Koike}, {Kurakami}, {Miyama}, {Morokuma}, {Nakata}, {Namikawa}, {Nakaya}, {Nariai}, {Obuchi}, {Oishi}, {Okada}, {Okura}, {Tait}, {Takata}, {Tanaka}, {Tanaka}, {Terai}, {Tomono}, {Uraguchi}, {Usuda}, {Utsumi}, {Yamada}, {Yamanoi}, {Aihara}, {Fujimori}, {Mineo}, {Miyatake}, {Oguri}, {Uchida}, {Tanaka}, {Yasuda}, {Takada}, {Murayama}, {Nishizawa}, {Sugiyama}, {Chiba}, {Futamase}, {Wang}, {Chen}, {Ho}, {Liaw}, {Chiu}, {Ho}, {Lai}, {Lee}, {Jeng}, {Iwamura}, {Armstrong}, {Bickerton}, {Bosch}, {Gunn}, {Lupton}, {Loomis}, {Price}, {Smith}, {Strauss}, {Turner}, {Suzuki}, {Miyazaki}, {Muramatsu}, {Yamamoto}, {Endo}, {Ezaki}, {Ito}, {Kawaguchi}, {Sofuku}, {Taniike}, {Akutsu}, {Dojo}, {Kasumi}, {Matsuda}, {Imoto}, {Miwa}, {Suzuki}, {Takeshi}, \& {Yokota}}]{Miyazaki2018}
{Miyazaki}, S., {Komiyama}, Y., {Kawanomoto}, S., {et~al.} 2018, \pasj, 70, S1

\bibitem[{{Momcheva} {et~al.}(2016){Momcheva}, {Brammer}, {van Dokkum}, {Skelton}, {Whitaker}, {Nelson}, {Fumagalli}, {Maseda}, {Leja}, {Franx}, {Rix}, {Bezanson}, {Da Cunha}, {Dickey}, {F{\"o}rster Schreiber}, {Illingworth}, {Kriek}, {Labb{\'e}}, {Ulf Lange}, {Lundgren}, {Magee}, {Marchesini}, {Oesch}, {Pacifici}, {Patel}, {Price}, {Tal}, {Wake}, {van der Wel}, \& {Wuyts}}]{Momcheva2016}
{Momcheva}, I.~G., {Brammer}, G.~B., {van Dokkum}, P.~G., {et~al.} 2016, \apjs, 225, 27

\bibitem[{Nair \& Hinton(2010)}]{nair2010rectified}
Nair, V. \& Hinton, G.~E. 2010, in Proceedings of the 27th international conference on machine learning (ICML-10), 807--814

\bibitem[{{Newman} {et~al.}(2013){Newman}, {Cooper}, {Davis}, {Faber}, {Coil}, {Guhathakurta}, {Koo}, {Phillips}, {Conroy}, {Dutton}, {Finkbeiner}, {Gerke}, {Rosario}, {Weiner}, {Willmer}, {Yan}, {Harker}, {Kassin}, {Konidaris}, {Lai}, {Madgwick}, {Noeske}, {Wirth}, {Connolly}, {Kaiser}, {Kirby}, {Lemaux}, {Lin}, {Lotz}, {Luppino}, {Marinoni}, {Matthews}, {Metevier}, \& {Schiavon}}]{Newman2013}
{Newman}, J.~A., {Cooper}, M.~C., {Davis}, M., {et~al.} 2013, \apjs, 208, 5

\bibitem[{Ngiam {et~al.}(2011)Ngiam, Khosla, Kim, Nam, Lee, \& Ng}]{ngiam2011multimodal}
Ngiam, J., Khosla, A., Kim, M., {et~al.} 2011, in Proceedings of the 28th international conference on machine learning (ICML-11), 689--696

\bibitem[{{Pasquet} {et~al.}(2019){Pasquet}, {Bertin}, {Treyer}, {Arnouts}, \& {Fouchez}}]{Pasquet2019}
{Pasquet}, J., {Bertin}, E., {Treyer}, M., {Arnouts}, S., \& {Fouchez}, D. 2019, \aap, 621, A26

\bibitem[{Qian {et~al.}(2021)Qian, Zhu, Zhang, \& Li}]{qian2021robust}
Qian, K., Zhu, S., Zhang, X., \& Li, L.~E. 2021, in Proceedings of the IEEE/CVF Conference on Computer Vision and Pattern Recognition, 444--453

\bibitem[{{Regnault} {et~al.}(2009){Regnault}, {Conley}, {Guy}, {Sullivan}, {Cuillandre}, {Astier}, {Balland}, {Basa}, {Carlberg}, {Fouchez}, {Hardin}, {Hook}, {Howell}, {Pain}, {Perrett}, \& {Pritchet}}]{Regnault2009}
{Regnault}, N., {Conley}, A., {Guy}, J., {et~al.} 2009, \aap, 506, 999

\bibitem[{Rogez {et~al.}(2017)Rogez, Weinzaepfel, \& Schmid}]{rogez2017lcr}
Rogez, G., Weinzaepfel, P., \& Schmid, C. 2017, in Proceedings of the IEEE Conference on Computer Vision and Pattern Recognition, 3433--3441

\bibitem[{Rothe {et~al.}(2018)Rothe, Timofte, \& Van~Gool}]{rothe2018deep}
Rothe, R., Timofte, R., \& Van~Gool, L. 2018, International Journal of Computer Vision, 126, 144

\bibitem[{{Sawicki} {et~al.}(2019){Sawicki}, {Arnouts}, {Huang}, {Coupon}, {Golob}, {Gwyn}, {Foucaud}, {Moutard}, {Iwata}, {Liu}, {Chen}, {Desprez}, {Harikane}, {Ono}, {Strauss}, {Tanaka}, {Thibert}, {Balogh}, {Bundy}, {Chapman}, {Gunn}, {Hsieh}, {Ilbert}, {Jing}, {LeF{\`e}vre}, {Li}, {Matsuda}, {Miyazaki}, {Nagao}, {Nishizawa}, {Ouchi}, {Shimasaku}, {Silverman}, {de la Torre}, {Tresse}, {Wang}, {Willott}, {Yamada}, {Yang}, \& {Yee}}]{Sawicki2019}
{Sawicki}, M., {Arnouts}, S., {Huang}, J., {et~al.} 2019, \mnras, 489, 5202

\bibitem[{{Schlegel} {et~al.}(1998){Schlegel}, {Finkbeiner}, \& {Davis}}]{Schlegel1998}
{Schlegel}, D.~J., {Finkbeiner}, D.~P., \& {Davis}, M. 1998, \apj, 500, 525

\bibitem[{{Schuldt} {et~al.}(2021){Schuldt}, {Suyu}, {Ca{\~n}ameras}, {Taubenberger}, {Meinhardt}, {Leal-Taix{\'e}}, \& {Hsieh}}]{Schuldt2021}
{Schuldt}, S., {Suyu}, S.~H., {Ca{\~n}ameras}, R., {et~al.} 2021, \aap, 651, A55

\bibitem[{{Scodeggio} {et~al.}(2018){Scodeggio}, {Guzzo}, {Garilli}, {Granett}, {Bolzonella}, {de la Torre}, {Abbas}, {Adami}, {Arnouts}, {Bottini}, {Cappi}, {Coupon}, {Cucciati}, {Davidzon}, {Franzetti}, {Fritz}, {Iovino}, {Krywult}, {Le Brun}, {Le F{\`e}vre}, {Maccagni}, {Ma{\l}ek}, {Marchetti}, {Marulli}, {Polletta}, {Pollo}, {Tasca}, {Tojeiro}, {Vergani}, {Zanichelli}, {Bel}, {Branchini}, {De Lucia}, {Ilbert}, {McCracken}, {Moutard}, {Peacock}, {Zamorani}, {Burden}, {Fumana}, {Jullo}, {Marinoni}, {Mellier}, {Moscardini}, \& {Percival}}]{Scodeggio2018}
{Scodeggio}, M., {Guzzo}, L., {Garilli}, B., {et~al.} 2018, \aap, 609, A84

\bibitem[{{Skelton} {et~al.}(2014){Skelton}, {Whitaker}, {Momcheva}, {Brammer}, {van Dokkum}, {Labb{\'e}}, {Franx}, {van der Wel}, {Bezanson}, {Da Cunha}, {Fumagalli}, {F{\"o}rster Schreiber}, {Kriek}, {Leja}, {Lundgren}, {Magee}, {Marchesini}, {Maseda}, {Nelson}, {Oesch}, {Pacifici}, {Patel}, {Price}, {Rix}, {Tal}, {Wake}, \& {Wuyts}}]{Skelton2014}
{Skelton}, R.~E., {Whitaker}, K.~E., {Momcheva}, I.~G., {et~al.} 2014, \apjs, 214, 24

\bibitem[{St{\"o}ter {et~al.}(2018)St{\"o}ter, Chakrabarty, Edler, \& Habets}]{stoter2018classification}
St{\"o}ter, F.-R., Chakrabarty, S., Edler, B., \& Habets, E.~A. 2018, in 2018 IEEE International Conference on Acoustics, Speech and Signal Processing (ICASSP), IEEE, 436--440

\bibitem[{{Szalay} {et~al.}(1999){Szalay}, {Connolly}, \& {Szokoly}}]{Szalay1999}
{Szalay}, A.~S., {Connolly}, A.~J., \& {Szokoly}, G.~P. 1999, \aj, 117, 68

\bibitem[{Szegedy {et~al.}(2015)Szegedy, Liu, Jia, Sermanet, Reed, Anguelov, Erhan, Vanhoucke, \& Rabinovich}]{szegedy2015going}
Szegedy, C., Liu, W., Jia, Y., {et~al.} 2015, in Proceedings of the IEEE conference on computer vision and pattern recognition, 1--9

\bibitem[{Treyer {et~al.}(2023)Treyer, Ait-Ouahmed, Pasquet, Arnouts, Bertin, \& Fouchez}]{treyer2023}
Treyer, M., Ait-Ouahmed, R., Pasquet, J., {et~al.} 2023, Monthly Notices of the Royal Astronomical Society, submitted

\bibitem[{{Weaver} {et~al.}(2022){Weaver}, {Kauffmann}, {Ilbert}, {McCracken}, {Moneti}, {Toft}, {Brammer}, {Shuntov}, {Davidzon}, {Hsieh}, {Laigle}, {Anastasiou}, {Jespersen}, {Vinther}, {Capak}, {Casey}, {McPartland}, {Milvang-Jensen}, {Mobasher}, {Sanders}, {Zalesky}, {Arnouts}, {Aussel}, {Dunlop}, {Faisst}, {Franx}, {Furtak}, {Fynbo}, {Gould}, {Greve}, {Gwyn}, {Kartaltepe}, {Kashino}, {Koekemoer}, {Kokorev}, {Le F{\`e}vre}, {Lilly}, {Masters}, {Magdis}, {Mehta}, {Peng}, {Riechers}, {Salvato}, {Sawicki}, {Scarlata}, {Scoville}, {Shirley}, {Silverman}, {Sneppen}, {Smolc̆i{\'c}}, {Steinhardt}, {Stern}, {Tanaka}, {Taniguchi}, {Teplitz}, {Vaccari}, {Wang}, \& {Zamorani}}]{Weaver2022}
{Weaver}, J.~R., {Kauffmann}, O.~B., {Ilbert}, O., {et~al.} 2022, \apjs, 258, 11

\end{thebibliography}

\begin{appendix}
\section{Classification or regression}
\label{App:ClassvsReg}
To study the impact of both classification and regression training strategies, we test different models on both the HSC 9 band and the CFHTLS datasets using the single modality scheme.

We test training the model with the regression module using three different losses : 
\begin{itemize}
\setlength\itemsep{1em}
    \item Root mean square-error (RMSE) : 
    
     $ RMSE = \sqrt{mean((z_{pred} - z_{spec})^2)} $

    \item  mean absolute error (MAE) :
    
    $ MAE = mean(|z_{pred} - z_{spec}|) $

    \item  normalized MAE (NMAE, with the residuals normalized by the value of the label) : 
    
    $NMAE = mean(\frac{|z_{pred} - z_{spec}|}{z_{spec}+1}) $
    
\end{itemize}

We test also a model trained solely  with classification, and one aided by a MAE regression, by combining with equal weight the two loss functions in the training. 

The results of these experiments are presented in  Table.~\ref{table:classif_vs_reg_table} .

 For the two datasets, the performances with the regression appear to depend on the choice of the loss function, with the normalized MAE leading to the best performances.  Overall the classification-based models outperform the regression ones (especially for the MAD) for both datasets independently of the depth and number of available bands.  It is even slightly improved when the classification is co-optimized with a regression for the HSC dataset (we use the classification module estimation in this case). \\

As a conclusion, we adopt a classification model aided by a regression for all the experiments presented in this work.

\begin{table}[ht!]
\def\arraystretch{1.3}%
\begin{tabular}{|c|c|c|c|}
\hline
Experiences &  $\sigma$  & $\eta$ & $<\Delta z>$ \\
 &  \num{e-3} & \% & $10^{-3}$ \\
\hline
\multicolumn{4}{|c|}{HSC 9 bands }             \\ \hline
 Classification and Regression MAE & \bf{08.36} & \bf{1.24}  & \bf{0.68}           \\ \hline
 Classification  & 08.66 & 1.33  & 1.20           \\ \hline
 Regression RMSE & 18.99 & 1.33  & 1.86           \\ \hline
 Regression MAE  & 13.04 & 1.26  & 1.57         \\ \hline
 Normalized Regression MAE  & 12.03 & 1.25  & 1.15          \\ \hline

\multicolumn{4}{|c|}{CFHTLS  }             \\ \hline
 Classification and Regression MAE & \bf{16.28} & 0.99  & 1.43          \\ \hline
 Classification  & \bf{16.28} & \bf{0.98}  & 1.46           \\ \hline
 Regression RMSE & 20.79 & 0.96  & 1.18           \\ \hline
 Regression MAE  & 18.26 & 0.99  & 1.10         \\ \hline
 Normalized Regression MAE  & 17.95 & 0.99  & \bf{-0.56}          \\ \hline
\end{tabular}
\caption{Global performances of classification and regression based models (see text) for the HSC 9 bands and CFHTLS dataset.  
}
\label{table:classif_vs_reg_table}
\end{table}

\section{Multimodality impact on outlier fraction and bias}
\begin{figure*}
    \centering
    \begin{subfigure}[b]{0.45\textwidth}
        \centering
        \includegraphics[width=\textwidth]{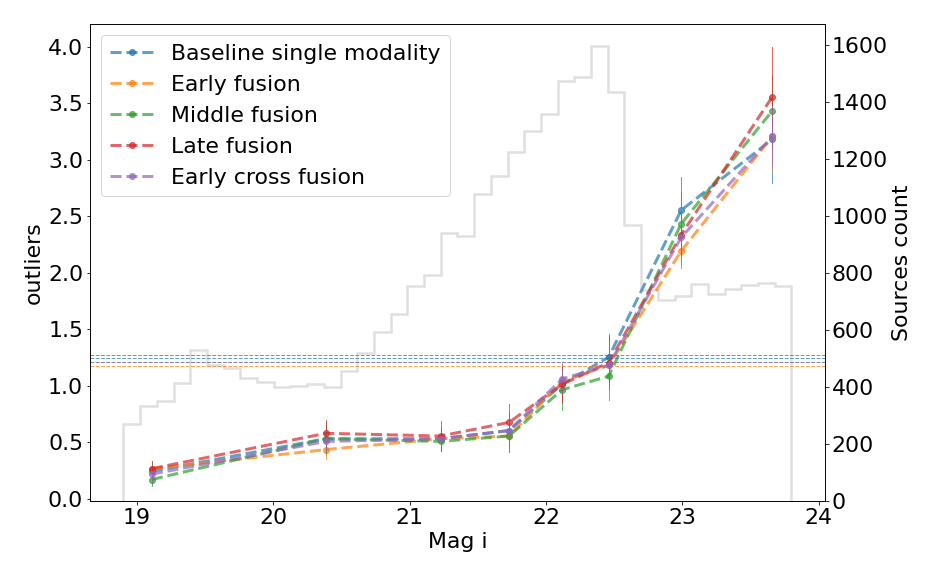}
        
    \end{subfigure}
    \hspace{\fill}
    \begin{subfigure}[b]{0.45\textwidth}
        \centering
        \includegraphics[width=\textwidth]{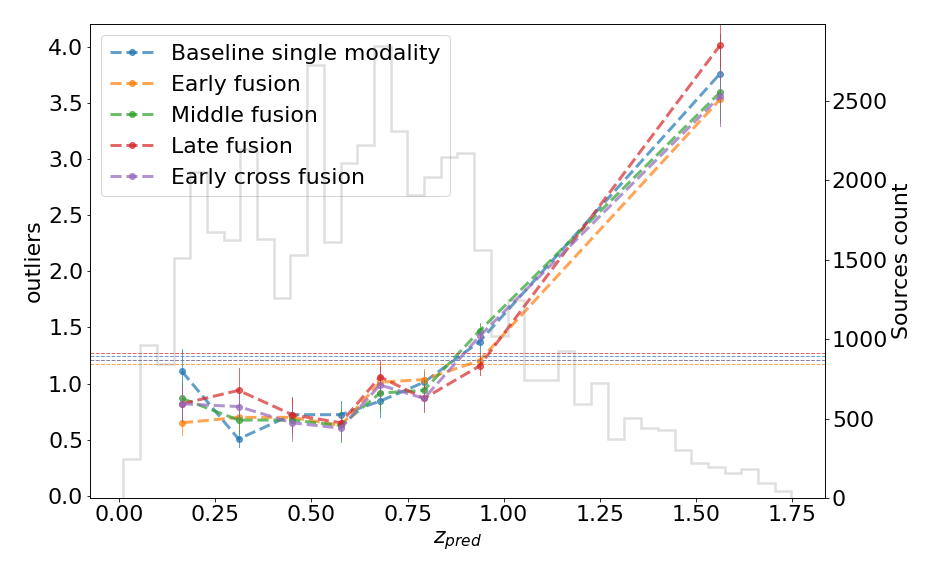}
        
    \end{subfigure}
    \\
    \begin{subfigure}[b]{0.45\textwidth}
        \centering
        \includegraphics[width=\textwidth]{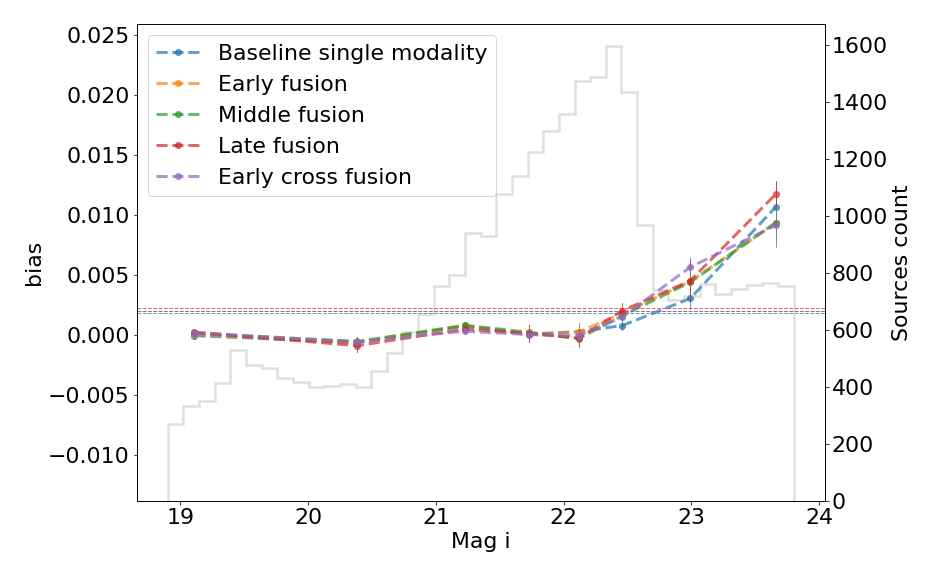}
        
    \end{subfigure}
    \hspace{\fill}
    \begin{subfigure}[b]{0.45\textwidth}
        \centering
        \includegraphics[width=\textwidth]{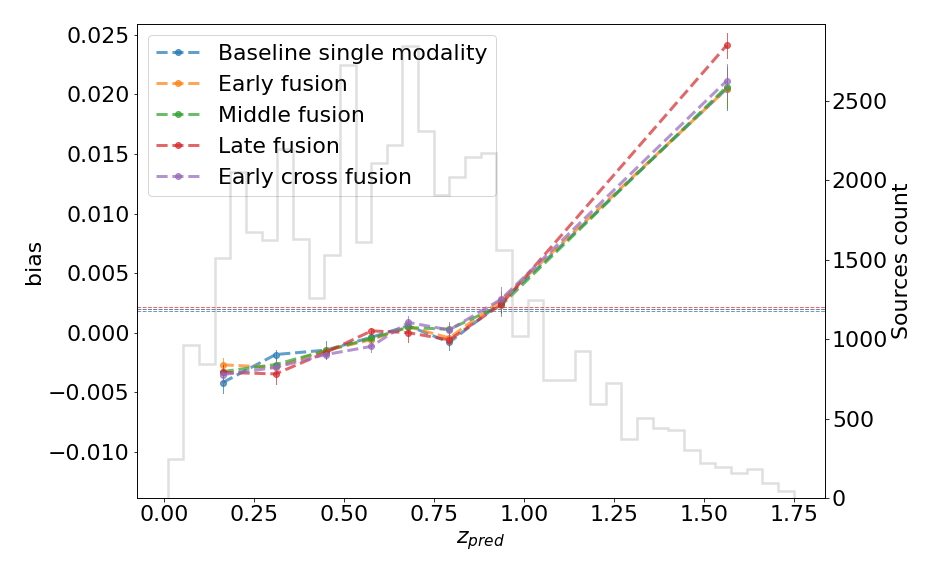}
        
    \end{subfigure}
        \caption{Comparison of the outlier fraction and bias versus band i magnitude  and predicted redshift  for different fusion types and the single modality baseline on the 9-band sources from the HSC dataset}
    \label{fig:o_and_b_fusion_type}
\end{figure*}

\begin{figure*}
    \centering
    \begin{subfigure}[b]{0.45\textwidth}
        \centering
        \includegraphics[width=\textwidth]{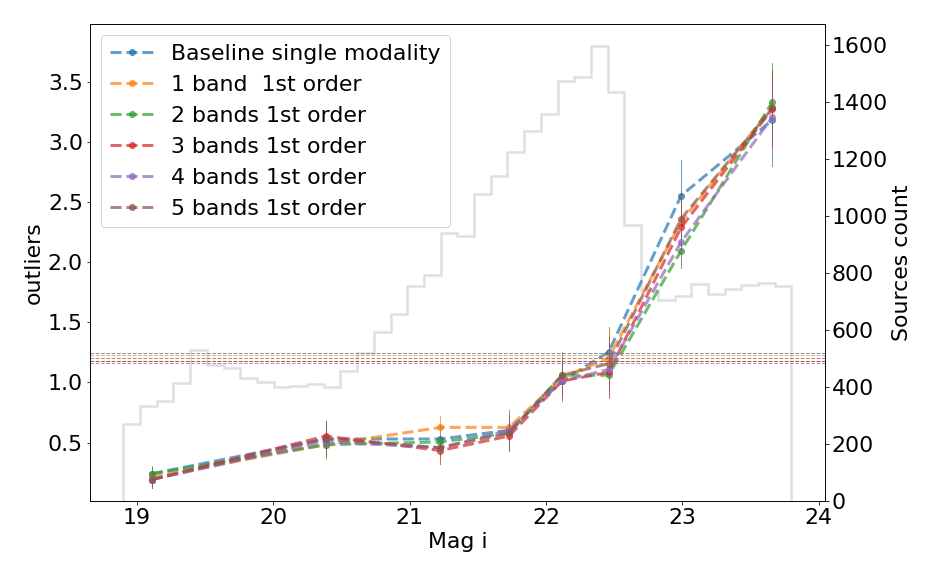}
        
    \end{subfigure}
    \hspace{\fill}
    \begin{subfigure}[b]{0.45\textwidth}
        \centering
        \includegraphics[width=\textwidth]{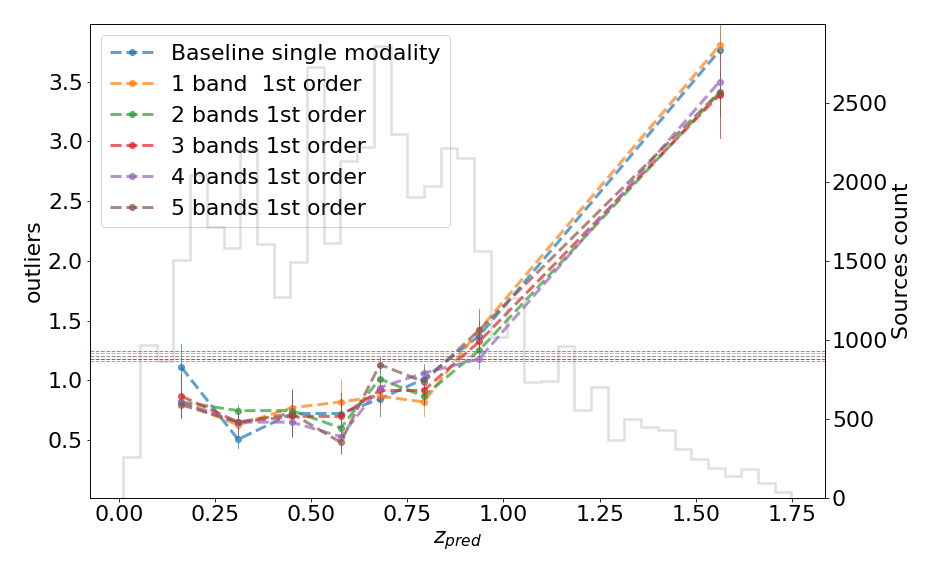}
        
    \end{subfigure}
    \\
    \begin{subfigure}[b]{0.45\textwidth}
        \centering
        \includegraphics[width=\textwidth]{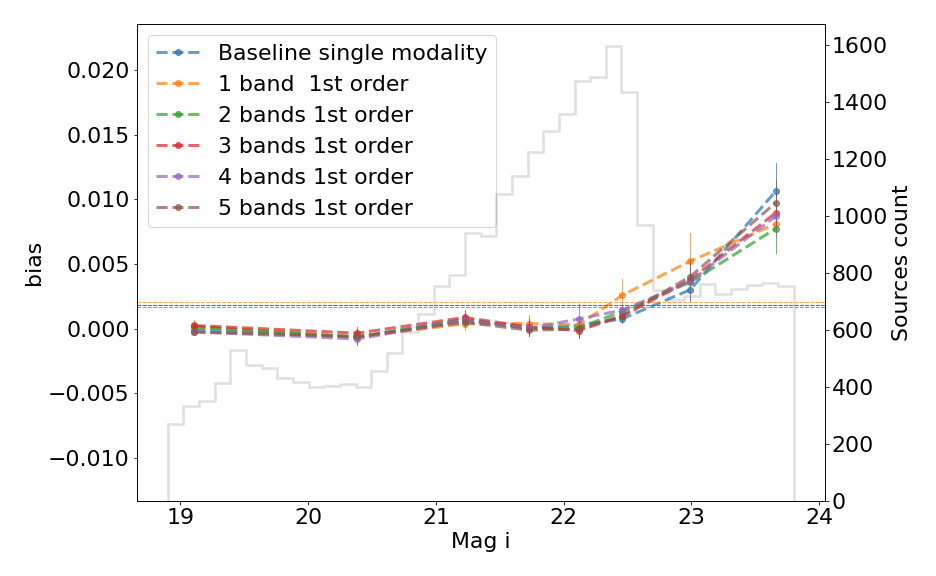}
        
    \end{subfigure}
    \hspace{\fill}
    \begin{subfigure}[b]{0.45\textwidth}
        \centering
        \includegraphics[width=\textwidth]{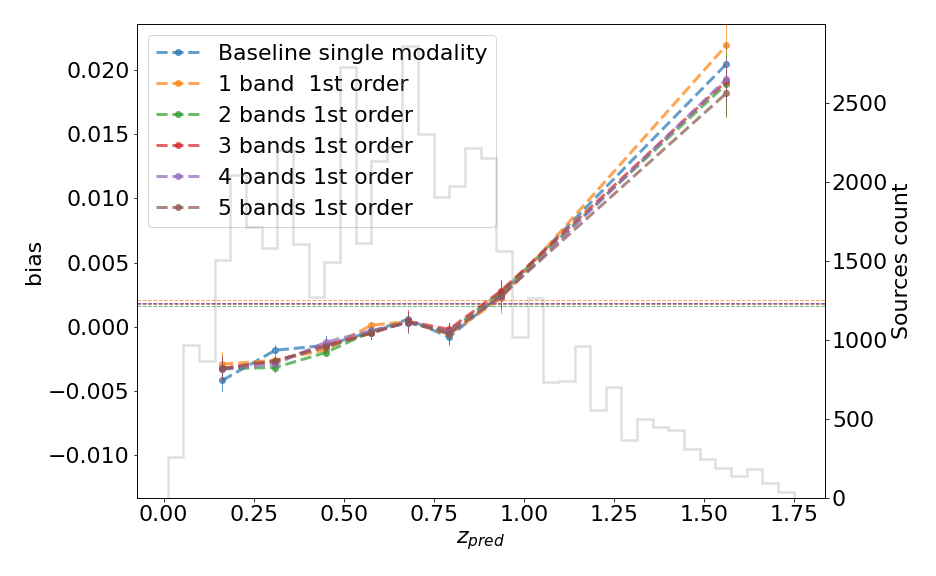}
        
    \end{subfigure}
    \caption{same as Fig ~\ref{fig:o_and_b_fusion_type}for early fusion first order modalities models with different number of bands per modality and the single modality baseline.}
    \label{fig:o_and_b_n_bands_per_modality}
\end{figure*}

\begin{figure*}
    \centering
    \begin{subfigure}[b]{0.45\textwidth}
        \centering
        \includegraphics[width=\textwidth]{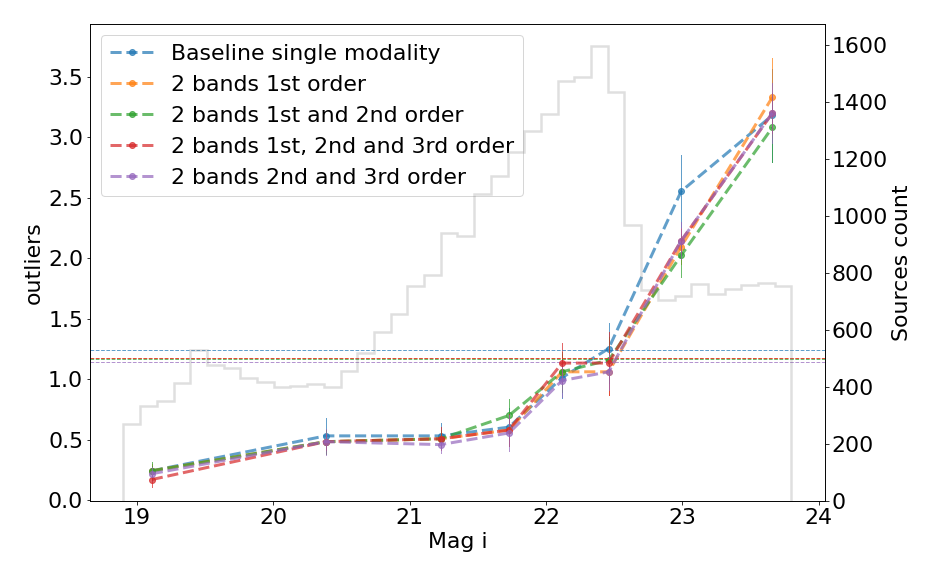}
        
    \end{subfigure}
    \hspace{\fill}
    \begin{subfigure}[b]{0.45\textwidth}
        \centering
        \includegraphics[width=\textwidth]{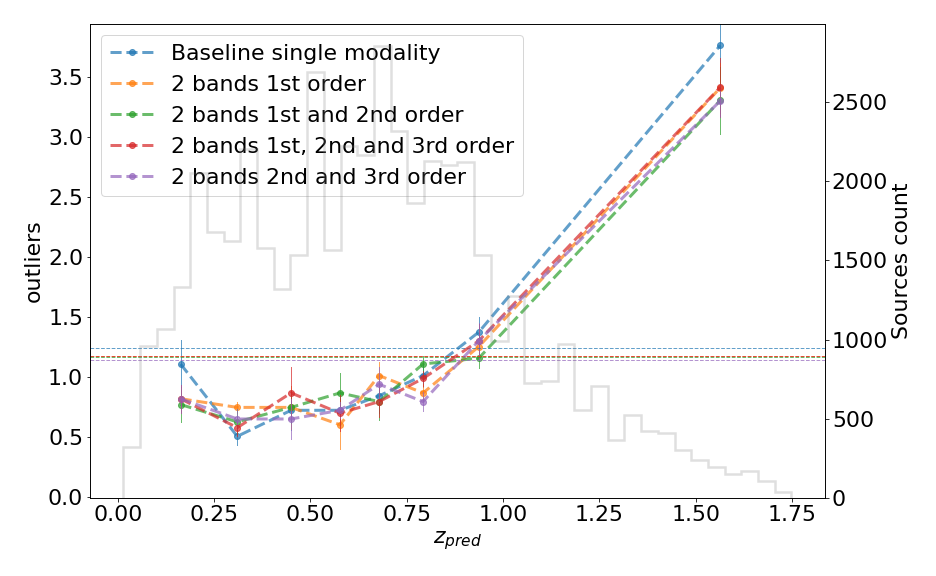}
        
    \end{subfigure}
    \\
    \begin{subfigure}[b]{0.45\textwidth}
        \centering
        \includegraphics[width=\textwidth]{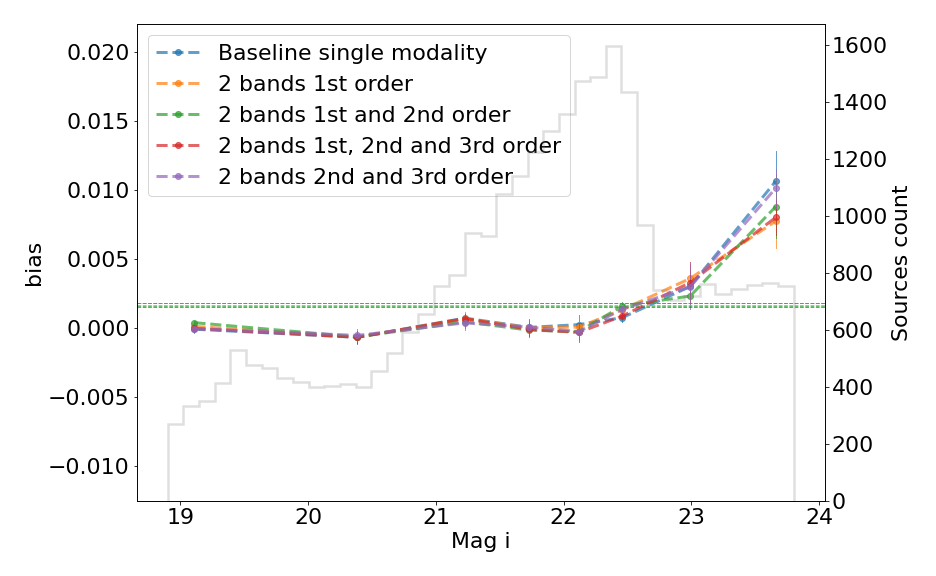}
        
    \end{subfigure}
    \hspace{\fill}
    \begin{subfigure}[b]{0.45\textwidth}
        \centering
        \includegraphics[width=\textwidth]{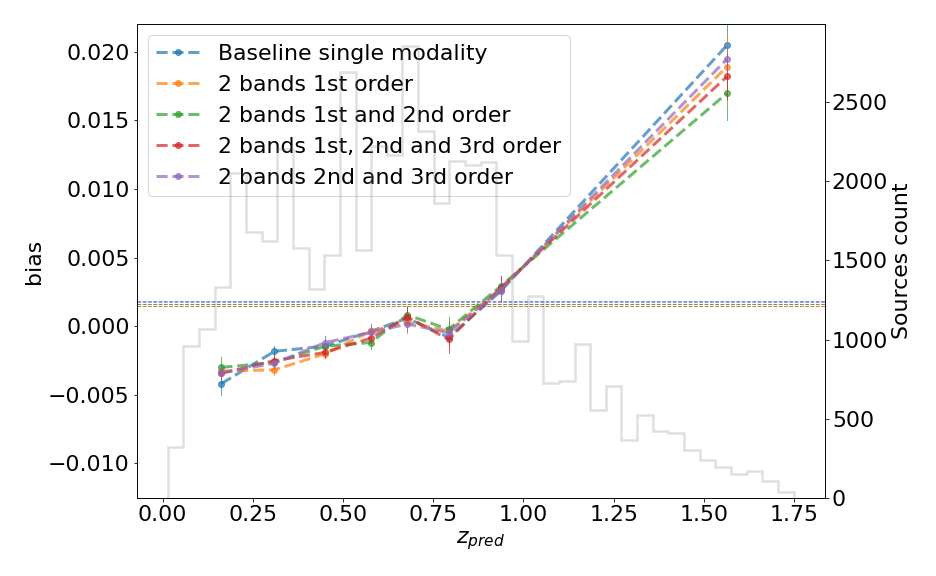}
        
    \end{subfigure}
    \caption{Same as Fig~\ref{fig:o_and_b_fusion_type} for different modality order combinations for 2 band modalities using early fusion and the single modality baseline. }
    \label{fig:o_and_b modality's order}
\end{figure*}

We previously detailed, for the different multimodality experiences, their impact on the MAD metric. Here we show the evolution of the outlier fraction and bias as a function of the i band magnitude and the estimated redshift $z_{pred}$.
We can see in the Figures ~\ref{fig:o_and_b_fusion_type}, ~\ref{fig:o_and_b_n_bands_per_modality}, ~\ref{fig:o_and_b modality's order} that different multimodality configurations slightly improve outlier fraction and have little impact compared to the baseline model on the bias. 

\begin{figure*}
    \centering
    \begin{subfigure}[b]{0.45\textwidth}
        \centering
        \includegraphics[width=\textwidth]{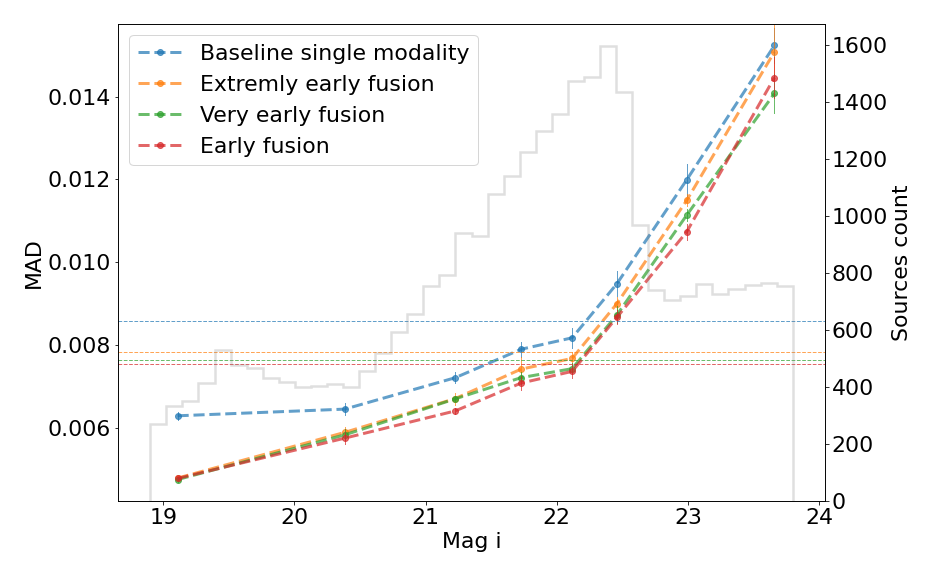}
        
    \end{subfigure}
    \hspace{\fill}
    \begin{subfigure}[b]{0.45\textwidth}
        \centering
        \includegraphics[width=\textwidth]{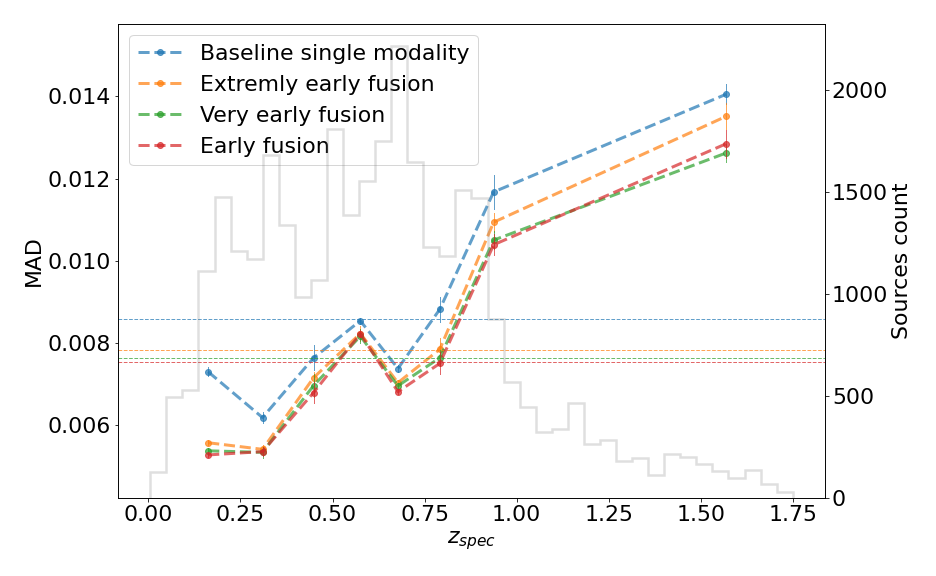}
        
    \end{subfigure}
     \caption{
     %Comparison of the redshift estimation metrics versus magnitude of band i and spectroscopic redshift 
     Same as Fig~\ref{fig:fusion} for early, very early and extremely early fusion models and the baseline model. % on the 9-band sources from the HSC dataset
     }
\label{fig:early_fusion_types}
\end{figure*}
\section{Configuration details}
\label{appendix:configuration_details}
% We layout the different details of the different configurations used through out this work. 
Table~\ref{table:models_n_parameters} shows the number of parameters for various experiments. We can note that certain multimodal models outperformed the baseline while having a lower number of parameters like the 2-band first-order modalities with early fusion. Additionally, other models with a higher number of parameters, such as the 5-band first-order modalities with early fusion, also showed improved performance. These results reinforce our conclusion that the effectiveness of the multimodal approach relies not on the number of parameters but rather on its superior capacity of relevant information extraction.

Table~\ref{table:modalities_order_details} details the composition of the modalities of first, second and third order and sizes of two, three and four bands per modality when using the \textit{ugrizyjhk} bands.

\begin{table}[ht!]
\begin{tabular}{|lc|}
\hline
\multicolumn{1}{|l|}{Experience}             & Parameters $(10^6)$ \\ \hline
\multicolumn{1}{|l|}{Baseline}               & 19.03               \\ \hline
\multicolumn{2}{|c|}{2 bands first order}                     \\ \hline
\multicolumn{1}{|l|}{Extremely early fusion}       & 9.52               \\ \hline
\multicolumn{1}{|l|}{Very early fusion}          & 11.16               \\ \hline
\multicolumn{1}{|l|}{Early fusion}                & 12.51               \\ \hline
\multicolumn{1}{|l|}{Middle fusion}                & 15.22               \\ \hline
\multicolumn{1}{|l|}{Late fusion}                & 17.94               \\ \hline
\multicolumn{2}{|c|}{Early fusion first order}                     \\ \hline
\multicolumn{1}{|l|}{1 band}                 & 8.80                \\ \hline
\multicolumn{1}{|l|}{2 bands}                & 12.51               \\ \hline
\multicolumn{1}{|l|}{3 bands}                & 17.23               \\ \hline
\multicolumn{1}{|l|}{4 bands}                & 22.11               \\ \hline
\multicolumn{1}{|l|}{5 bands}                & 26.29               \\ \hline
\multicolumn{2}{|c|}{Early fusion 2 bands}                         \\ \hline
\multicolumn{1}{|l|}{1st and 2nd order}      & 17.37               \\ \hline
\multicolumn{1}{|l|}{1st, 2nd and 3rd order} & 21.54               \\ \hline
\multicolumn{1}{|l|}{2nd and 3rd  order}     & 15.98               \\ \hline
\end{tabular}
\caption{Number of trainable parameters for some models grouped by fusion stage, modality size and modalities' orders. }
\label{table:models_n_parameters}
\end{table}

\begin{table}[ht!]
\begin{tabular}{|l|c|c|clllll|}
\hline & 2 bands & 3 bands                                                                                            & \multicolumn{6}{c|}{4 bands}                                                                                                 \\ \hline
1st order  & \begin{tabular}[c]{@{}c@{}}u\_g, g\_r,\\ r\_i, i\_z,\\ z\_y, y\_j,\\ j\_h, h\_k\end{tabular} & \begin{tabular}[c]{@{}c@{}}u\_g\_r, g\_r\_i,\\ r\_i\_z, i\_z\_y,\\ z\_y\_j, y\_j\_h,\\ j\_h\_k\end{tabular} & \multicolumn{6}{c|}{\begin{tabular}[c]{@{}c@{}}u\_g\_r\_i, g\_r\_i\_z,\\ r\_i\_z\_y, i\_z\_y\_j,\\ z\_y\_j\_h, y\_j\_h\_k\end{tabular}} \\ \hline
2nd order & \begin{tabular}[c]{@{}c@{}}u\_r, g\_i,\\ r\_z, i\_y,\\ z\_j, y\_h,\\ j\_k\end{tabular}       & \begin{tabular}[c]{@{}c@{}}u\_r\_z, g\_i\_y,\\ r\_z\_j, i\_y\_h,\\ z\_j\_k{]}\end{tabular}                  & \multicolumn{6}{c|}{\begin{tabular}[c]{@{}c@{}}u\_r\_z\_j, g\_i\_y\_h,\\ r\_z\_j\_k\end{tabular}}                                       \\ \hline
3rd order  & \begin{tabular}[c]{@{}c@{}}u\_i, g\_z,\\ r\_y, i\_j,\\ z\_h, y\_k\end{tabular}               & \begin{tabular}[c]{@{}c@{}}u\_i\_j, g\_z\_h,\\ r\_y\_k\end{tabular}                                         & \multicolumn{6}{c|}{}                                                                                                                   \\ \hline
\end{tabular}
\caption{A detailed breakdown of the specific bands that are included in each modality for all the first, second, and third order modalities of sizes 2, 3, and 4 in a 9 band dataset.}
\label{table:modalities_order_details}
\end{table}

\section{Paired bootstrap test}
\label{paired_bootstrap_test}
To assess the statistical significance of the observed difference between the baseline and the multimodal approaches, we use the paired bootstrap significance test introduced by \cite{efron1994introduction} and frequently used in the field of natural language processing \citep{berg-kirkpatrick-etal-2012-empirical, koehn-2004-statistical}. It is a non-parametric hypothesis test with no assumption about the distribution of the data. For a given dataset, $D$, we define: 
\begin{equation}
    \delta(D) = M_M(D)-M_B(D) 
\end{equation}
where $M_M(D)$ and $M_B(D)$ are the metrics of the multimodal and the baseline model respectively for the dataset $D$.
We first assume that $M_M$ is, contrary to what we believe, equal or worse than $M_B$. This is known as the null hypothesis, $H_0$. Next, for a given dataset $D_{test}$, we estimate the likelihood, $p_{value}(D_{test})$, of observing, under $H_0$ and on a new dataset $D$, a metric gain $\delta M(D)$ equal to or better than $\delta M(D_{test})$, so that :

\begin{equation}
p_{value}(D_{test}) = P(\delta(D) \ge \delta(D_{test})|H_0) 
\end{equation}

A low $p_{value}(D_{test})$ suggests that observing $\delta(D_{test})$ is unlikely if $H_0$ was true, so we can reject $H_0$ and conclude that the metric gain $\delta(D_{test})$ of $M_M$ compared to $M_B$ is significant and not just a random fluke.

The $p_{value}(D_{test})$ is hard to compute and must be  approximated as we don't have new data sets to test on, so we use the paired bootstrap method to simulate this. We sample from the test set, with replacement, $K$ same size samples as the test set, on which $\delta(D_{test})$ was computed. We refer to these samples as bootstrapped samples.

Naively we may think that we should compute the frequency of $\delta(D_{bootstrapped}) \ge \delta(D_{test})$ over the $K$ samples as an approximation of  $p_{value}(D_{test})$.
However, these samples won't be suitable for our null hypothesis $H_0$ since they were sampled from the test set, causing their average  $\delta(D_{bootstrapped})$ to be around $\delta(D_{test})$ contrary to what \(H_0\) requires. 
% The solution is to re-center $\delta(D_{bootstrapped})$ by subtracting  $\delta(D_{test})$:  
Because $H_0$ assumes that the initially observed difference $\delta(D_{test})$ is due to a random fluke, the solution is to shift the $\delta(D_{bootstrapped})$ distribution by this value, so we obtain:
\begin{equation}
p_{value}(D_{test}) = Freq(\delta(D_{bootstrapped}) - \delta(D_{test}) \ge \delta(D_{test}))
\end{equation}

The results reported in Table.~\ref{table:HSC} are obtained with a  significance test assuming \(K = 10^{4}\).

\begin{figure*}
\centering
\begin{subfigure}{0.48\textwidth}
\centering
\includegraphics[width=\linewidth]{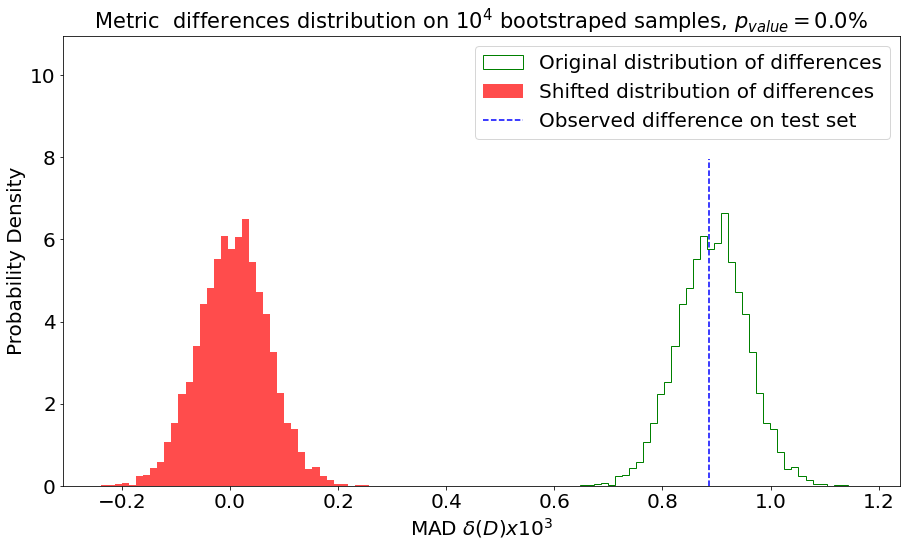}

\end{subfigure}%
\vspace{0.01\textwidth}
\begin{subfigure}{0.48\textwidth}
\centering
\includegraphics[width=\linewidth]{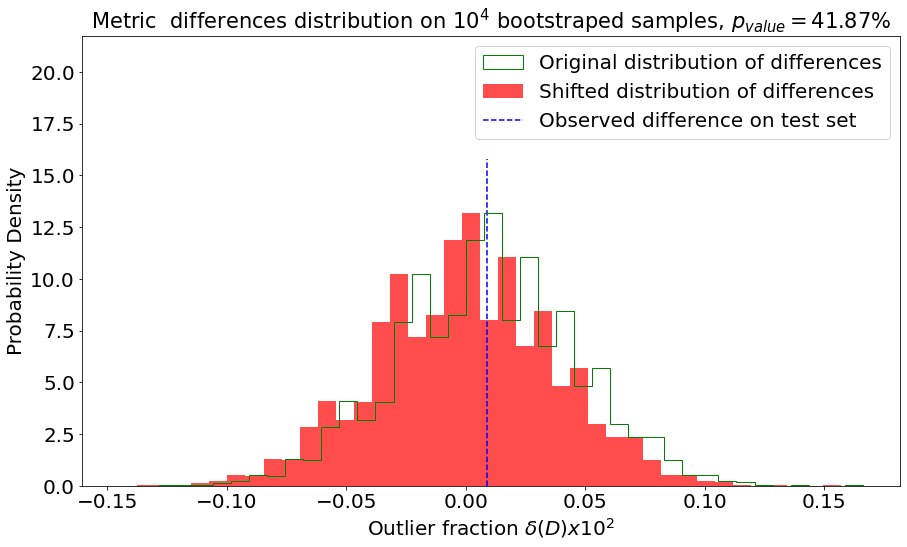}
\end{subfigure}%
\caption{Distribution of the MAD  and outlier fraction differences for a \(K = 10^4\) paired bootstrap test to asses the significance of the difference between $M_M$ and $M_B$ on the HSC 9 band test dataset.  }
\label{fig:bootstrap_test}
\end{figure*}

Fig.~\ref{fig:bootstrap_test} illustrates two cases for the HSC 9 band dataset: the left panel shows the  distributions for the  MAD of the bootstrapped samples, where the performance gain is significant; the right panel shows the distributions for the outlier fractions where the gain is not significant under a 5\% risk threshold. 
The green histogram is the original distribution that does not satisfy $H_0$, the red histogram is the shifted one that satisfies $H_0$. The blue line represents the difference initially observed on the test dataset $\delta(D_{test})$, and the $p_{value}(D_{test})$ corresponds to the fraction of the red histogram exceeding this value. 

\end{appendix}
%%%%%%%%%%%%%%%%%%%%%%%%%%%%%%%%%%%%%%%%%%%%%%%%%%
\end{document}